\documentclass[12pt,a4paper]{article}
\pdfoutput=1
\usepackage{jheppub}
\usepackage{latexsym}
\usepackage[utf8]{inputenc}
\usepackage{titlesec}
\usepackage{hyperref}
\usepackage[normalem]{ulem}
\usepackage{graphicx}  
\usepackage{amsmath,hepunits}
\usepackage{bm}  
\usepackage{xcolor}
\usepackage{amsfonts,amsmath,amssymb,bm,bbm}

\hypersetup{
    pdfnewwindow=true,      
    colorlinks=true,       
    linkcolor=blue,          
    citecolor=blue,        
    filecolor=blue,      
    urlcolor=blue        
}

\newcommand{\be}{\begin{equation}}
\newcommand{\ee}{\end{equation}}
\newcommand{\beq}{\begin{equation}}
\newcommand{\eeq}{\end{equation}}
\newcommand{\bqa}{\begin{eqnarray}}
\newcommand{\eqa}{\end{eqnarray}}
\newcommand{\bse}{\begin{subequations}}
\newcommand{\ese}{\end{subequations}}

\newcommand{\eq}[1]{Eq.~\eqref{eq:#1}}

\newcommand{\eqss}[3]{Eqs.~\eqref{eq:#1}, \eqref{eq:#2}, and \eqref{eq:#3}}

\begin{document}

\preprint{INT-PUB-18-027, MITP/18-036}

\title{Production of dark-matter bound states\\ in the early universe\\ by three-body recombination}

\author[a]{Eric Braaten,}
\emailAdd{braaten.1@osu.edu}
\author[b]{Daekyoung Kang,}
\emailAdd{dkang@fudan.edu.cn}
\author[c]{and Ranjan Laha}
\emailAdd{ranjalah@uni-mainz.de}

\affiliation[a]{Department of Physics,
         The Ohio State University, Columbus, OH\ 43210, USA}

\affiliation[b]{Key Laboratory of Nuclear Physics and Ion-beam Application (MOE) and \\ 
Institute of Modern Physics, 
Fudan University, Shanghai, China 200433}

\affiliation[c]{PRISMA Cluster of Excellence and
             Mainz Institute for Theoretical Physics, \\
             Johannes Gutenberg-Universit\"{a}t Mainz, 55099 Mainz, Germany}

\date{\today}

\abstract{The small-scale structure problems of the universe can be solved by self-interacting dark matter
that becomes strongly interacting at low energy.
A particularly predictive model for the self-interactions is resonant short-range interactions with
an S-wave scattering length that is much larger than the range.  
The velocity dependence of the cross section in such a model provides an excellent fit to 
self-interaction cross sections inferred from dark-matter halos of galaxies and clusters of galaxies
if the dark-matter mass is about 19~GeV and the scattering length is about 17~fm.
Such a model makes definite predictions for the few-body physics of weakly bound clusters of the dark-matter particles.
The formation of the two-body bound cluster is a bottleneck for the formation of larger bound clusters.
We calculate the production of two-body bound clusters by three-body recombination in the early universe
under the assumption that the dark matter particles are identical bosons, which is the most favorable case.
If the dark-matter mass is  19~GeV and the scattering length is 17~fm,
 the fraction of dark matter in the form of two-body bound clusters 
can increase by as much as 4 orders of magnitude
when the dark-matter temperature falls below the binding energy, but its  present value remains less than $10^{-6}$.
The present  fraction can be increased to as large as $10^{-3}$ by relaxing the constraints from small-scale structure
and decreasing the mass of the dark matter particle. 
}

\keywords{
Dark matter, Bound State, Effective Field Theories, Beyond Standard Model}
\maketitle

\section{Introduction}
\label{sec:Intro}

The simplest paradigm for dark matter is that it consists of weakly interacting elementary particles.  However visible matter consists not only of elementary particles, such as electrons,
but also of composite particles, such as nuclei.  
Nuclei are clusters of protons and neutrons bound by residual forces from QCD.  Protons and neutrons consist of quarks bound by the color force of QCD.
Dark matter could also consist of composite particles.  
In particular, it could consist of ``dark nucleons" and bound clusters of dark nucleons (``dark nuclei'').
The dark nucleons could be elementary or composite, 
but they have an integer-valued conserved charge that we call ``dark baryon number''.
Various models for dark-matter bound states have been discussed in the literature,
 including a near-threshold S-wave resonance 
 \cite{Braaten:2013tza, Laha:2013gva, Laha:2015yoa}, 
 the exchange of a light mediating boson between elementary fermions, 
 QCD-like structure in the dark sector, and other mechanisms\,\cite{Shepherd:2009sa, Khlopov:2010pq, Cline:2013zca, Foot:2014mia, Petraki:2014uza, Detmold:2014qqa, Detmold:2014kba, Wise:2014jva, vonHarling:2014kha, Wise:2014ola, Hardy:2014mqa, Appelquist:2015yfa, Hardy:2015boa, Petraki:2015hla, An:2015pva,Tsai:2015ugz, Bi:2016gca, Kribs:2016cew,
Kouvaris:2016ltf, Nozzoli:2016coi, Butcher:2016hic, Asadi:2016ybp, Petraki:2016cnz, Cirelli:2016rnw, Lonsdale:2017mzg, Gresham:2017zqi, Mitridate:2017oky, Elor:2018xku, Harz:2018csl,Biondini:2018pwp,Biondini:2018xor}. 
Light nuclei up to $^7$Li are produced in the early universe by big-bang nucleosynthesis\,\cite{Cyburt:2015mya, Consiglio:2017pot}.
The relevant reactions are all 2-body collisions of nuclei. 
Some of the 2-body reactions that produce a nucleus with larger baryon number than either
of the colliding nuclei are rearrangement reactions, such as $d + d \to {}^3{\rm He} + n$.
However the most important such reactions are radiative fusion reactions, such as $p + d \to {}^3{\rm He} + \gamma$,
in which the two incoming nuclei coalesce while radiating a photon to conserve energy and momentum.  The effects of 3-body collisions are negligible in big-bang nucleosynthesis.
Three-body collisions do play a role in stellar nucleosynthesis despite the relatively low density.
In particular, the Hoyle reaction $\alpha + \alpha + \alpha \to {}^{12}{\rm C} + \gamma$
provides a pathway around the bottleneck caused 
by the relatively large binding energy of the ${}^4{\rm He}$ nucleus $\alpha$.

If dark matter consists of dark nucleons that can form bound clusters,
these dark nuclei can be produced in the early universe by ``dark nucleosynthesis''.  
Studies have shown that a sequence of dark nuclei with increasing
dark baryon number can indeed be produced in the early universe
\cite{Krnjaic:2014xza,Detmold:2014qqa,Wise:2014jva,Hardy:2014mqa}.
The relevant few-body mechanisms were assumed to be 2-body radiative fusion reactions,
in which two incoming dark nuclei coalesce while radiating a much lighter particle 
to conserve energy and momentum. 
If there is no such light particle, dark nuclei with larger dark baryon numbers
must instead be built up through rearrangement collisions.
If dark nuclei with dark baryon number 2 (``dark deuterons'') 
have already been formed,
dark nuclei with larger dark baryon numbers can be produced by rearrangement collisions
of two dark nuclei, in which dark nucleons are transferred between the two colliding nuclei.
However the production of the dark deuterons is a bottleneck
that can only be overcome by collisions of 3 or more dark nucleons.
The simplest such reaction is the 3-body recombination of three dark nucleons into
a dark deuteron and a recoiling dark nucleon.
Whether a significant population of dark deuterons can be produced in the early universe
can only be determined by detailed calculations using specific models for the few-body physics.

One class of models for few-body physics that are extremely predictive
are those with short-range interactions 
and an S-wave resonance very close to the scattering threshold
for a pair of particles \cite{Braaten:2004rn}.
In these models, the elastic scattering cross section for a pair of particles 
has dramatic energy dependence that is completely determined by the 
particle mass $m$ and the S-wave scattering length $a$,
which is much larger that the range $r_0$ of the interactions.
When the center-of-mass collision energy $E$ decreases below the energy scale $1/m r_0^2$  set by the range, 
the elastic cross section  increases  as $1/E$,
nearly saturating the S-wave unitarity bound.  The  cross section levels off when $E$
decreases below the energy scale $1/m a^2$,
approaching a large value proportional to  $a^2$ as $E$ approaches 0.
If $a$ is positive, the S-wave resonance is a stable bound cluster.
This weakly bound cluster is {\it universal}, in the sense that it has properties determined by $a$, including a
small binding energy $1/m a^2$ and large geometric size of order $a$.
The universality in the two-particle sector for particles with a large scattering length extends to the 3-particle 
and higher sectors, although it can be more intricate \cite{Braaten:2004rn}.
It strongly constrains the universal bound clusters, 
whose binding energies are smaller than the energy scale set by the range.
Whether there are universal bound clusters with 3 or more particles
depends on the symmetries of the particles.
The simplest case in which there are such universal bound clusters  is identical spin-0 bosons.  
Universality also provides strong constraints on reaction rates in the 3-particle 
and higher sectors \cite{Braaten:2004rn}.
These reaction rates can display dramatic resonant  enhancements at low energy.
For example, the 3-body recombination rate can increase as $1/E^2$ 
when the center-of-mass collision energy $E$ decreases below the energy scale $1/m r_0^2$,
and it can approach a large value proportional to $a^4$ in the low-energy limit.

Large low-energy cross sections for dark matter particles are motivated by discrepancies 
between observations of the small-scale structure of the universe 
and simulations based on collisionless cold dark matter\,\cite{Weinberg:2013aya, Brooks:2014qya, Pontzen:2014lma, DelPopolo:2016emo, Bullock:2017xww, Buckley:2017ijx}.
Observations of dwarf galaxies are inconsistent with the cusp of dark matter at the center
of a galaxy that is predicted by dark-matter-only simulations.  Dark-matter-only simulations also imply that dwarf galaxies
bound to the Milky Way should be denser 
than those that have been observed.
Although other explanations for these problems have been proposed, they can all be solved by 
self-interacting dark matter that is strongly interacting at low energy\,\cite{Spergel:1999mh, Tulin:2017ara}.
Short-range interactions with a large scattering length provide a particularly 
predictive model of self-interactions  that become strong at low energy\,\cite{Braaten:2013tza, Laha:2013gva, Laha:2015yoa}.

If dark nucleons have short-range interactions with a large scattering length $a$,
 they have universal low-energy properties determined by $a$ \cite{Braaten:2013tza}.
We denote the dark nucleon by $d$ and a bound cluster of $n$ dark nucleons  by $d_n$.
If $a$ is negative, there are no  weakly  bound clusters of two dark nucleons.
If $a$ is positive, there is one universal weakly  bound cluster: the dark deuteron $d_2$. 
If a pair of dark nucleons has annihilation channels, 
the scattering length $a$ is complex with a small negative imaginary part.
In addition to the elastic cross section and the binding energy of the dark deuteron,
 the annihilation rate of a pair of dark nucleons 
and the decay rate of the dark deuteron are also universal
in the sense that they are determined by the complex scattering length \cite{Braaten:2013tza}.
In a direct detection experiment, the dark deuteron can scatter elastically from a target nucleus,
or it can be broken up by the collision\,\cite{Laha:2013gva, Laha:2015yoa, Laha:2016iom}.  
The low-energy cross sections for both processes
are determined by $a$ up to a multiplicative factor.
Their dependence on the collision energy and the recoil angle provides interesting signatures 
for this simplest dark nucleus. 
The simplest reaction that can form a bound cluster is 
3-body recombination: $d + d + d  \to d_2 + d$.
In an expanding and cooling thermal system, such as the early universe,
the decreasing number density will tend to suppress 3-body recombination
while the decreasing temperature will tend to enhance it.
Once dark deuterons are produced,
the competing breakup reaction  $d_2 + d \to d + d + d$ will destroy them.  
The net effect on the population of $d_2$ can only be determined by explicit calculations.

In this paper, we study 3-body recombination into dark deuterons during the Hubble expansion in the early universe
under the assumption that the dark matter consists of dark nucleons that are identical bosons with a large positive scattering length,
which is the most favorable case for the formation of universal bound clusters.
We determine the mass $m_\chi$ and the scattering length $a$ of the dark nucleon
that would be required to solve the small-scale structure problems of the universe.
For these values of $m_\chi$ and  $a$,
the fraction of dark matter in the form of dark deuterons
can increase by orders of magnitude when the dark-matter temperature decreases
to below the binding energy of the dark deuteron.
However, we find that a significant population of dark deuterons cannot be produced during the Hubble expansion.
Since the production of the dark deuteron is a bottleneck, larger dark nuclei will also not be formed.
A much larger population of dark deuterons can be produced if the constraints from small-scale structure
are relaxed and  the mass of the dark matter particle is decreased.

In Section~\ref{sec:two-body},
we summarize the universal 2-body physics of particles with a large scattering length.
We also determine the mass and the scattering length of the dark nucleon
that would be required to solve small-scale structure problems of the universe.
In Section~\ref{sec:three-body},
we summarize the universal 3-body physics of identical bosons with a large scattering length.
In Section~\ref{sec:thermal},
we present results for the rate constants for many-body systems of identical bosons 
with a large scattering length in thermal equilibrium.
In Section~\ref{sec:universe}, we consider the formation of dark deuterons 
by 3-body recombination during the Hubble expansion of the early universe.
We calculate the fraction of dark matter in the form of  dark deuterons 
as a function of the red shift.
Our results are summarized and discussed in Section~\ref{sec:discussion}.

\section{Universal two-body physics with large scattering length}
\label{sec:two-body}

In this section, we summarize the universal two-body physics of  
particles with short-range self-interactions and a large scattering length.
We determine the mass and the large scattering length of a dark nucleon 
that would be required to solve the small-scale structure problems of the universe.

\subsection{Two-body physics}

Atomic physics has provided a strong impetus for  developing the universal few-body physics 
of particles with large scattering lengths  \cite{Braaten:2004rn}.
There are naturally occurring atoms with large scattering lengths, such as the ${}^4$He atom.
There are other atoms whose scattering lengths  can  be controlled and made arbitrarily large 
by using Feshbach resonances \cite{chin2010feshbach}.  In this subsection, 
we  use the concise language of atomic physics for the particles with large scattering lengths and their bound clusters.  
The particle $d$ is referred to as an {\it atom}, and
the  two-body bound cluster $d_2$ is called a {\it dimer}.
We make factors of Planck's constant $\hbar$ explicit.

We denote the mass of the atom $d$ by $m$.
The atom has short-range self-interactions with range $r_0$ 
and an S-wave scattering length $a$ that is much larger than $r_0$. 
The range and the scattering length provide a high energy scale $E_0 = \hbar^2/m r_0^2$
and a low energy scale $E_2=\hbar^2/ma^2$.
At energies well below $E_0$,
the two-body physics is universal in the sense that it is completely determined by $a$.
It depends on the nature of the particles
and on the details of their short-range interactions only through $a$.
The universal behavior becomes exact in the zero-range limit $r_0 \to 0$.
In this limit, all higher partial-wave interactions go to 0,
so two-body scattering is purely S-wave.

The universal region for the scattering of two atoms is
when the collision energy $E$,
which is the kinetic energy in the center-of-mass frame, is well below $E_0$.
The universal elastic scattering cross section for identical bosons is
\be
\sigma_\mathrm{elastic}(E) =\frac{8 \pi}{1/a^2 + m E/\hbar^2} .
\label{eq:sigma}
\ee
If the two colliding atoms are distinguishable particles,
such as the two spin states of a spin-$\frac12$ fermion, the  numerator is replaced by $4\pi$.
The cross section has dramatic energy dependence.
When the collision energy $E$ decreases below $E_0$, 
the elastic cross section  increases in accordance with Eq.~\eqref{eq:sigma}.
In the scaling region $E_2 \ll E \ll E_0$, the cross section nearly saturates the S-wave unitarity bound
$8 \pi \hbar^2/mE$.
As the energy decreases below $E_2$, the cross section levels off
and approaches its maximum value $8 \pi a^2$ as $E\to0$.
In the limit $a \to \pm \infty$, the scaling behavior $8 \pi \hbar^2/mE$
extends down to arbitrarily low energy.  
Since this cross section saturates the S-wave unitarity bound,
the limit $a \rightarrow \pm \infty$  is called the {\it unitary limit}.

A universal bound state is one that has properties determined by $a$.
Its  binding energy per pair of particles must be less than $E_0$.
Whether or not there is a universal dimer $d_2$ depends on the sign of $a$.
If $a<0$, there is no universal dimer.
If $a>0$, there is a single universal dimer.
The universal binding energy of $d_2$ in the zero-range limit is
\be
E_2 = \hbar^2/ma^2 .
\label{eq:E2}
\ee

A beautiful example in atomic physics of a boson with a large scattering length 
is the $^4$He atom.  Its scattering length is about 200~$a_0$,
where $a_0$ is the Bohr radius.  The scattering length  is larger 
than the effective range by about a factor of 15,
so the cross section increases at low energies by 
more than two orders of magnitude.  The $^4$He dimer
is a universal two-body bound state with the tiny binding energy 
$E_2=1.4 \times 10^{-7}$~eV.
The $^4$He dimer was first observed in 1993 using electron impact ionization \cite{luo1993weakest}.
The universal low-energy behavior of particles with a large scattering length
is illustrated even more dramatically by experiments
with ultracold trapped atoms.  
The scattering length $a$ of the atoms can be controlled 
and made arbitrarily large by tuning the magnetic field to a 
Feshbach resonance \cite{chin2010feshbach}.
Thus the binding energy of the universal dimer can be controlled 
and made arbitrarily small.

If the atoms have inelastic scattering channels, 
the scattering length $a$ is complex with a negative imaginary part.
If all the inelastic scattering channels have energy release large compared to $E_2$,
the inclusive inelastic cross section is also universal and determined by $a$.
The universal inelastic scattering cross section for identical bosons is
\be
\sigma_\mathrm{inelastic}(E) = 
\frac{8 \pi\, \mathrm{Im}[1/a] }{(m E/\hbar^2)^{1/2}  \big[ 1/a^2 + m E/\hbar^2\big]} .
\label{eq:sigmain}
\ee
We have assumed the imaginary part of $1/a$ is tiny compared to the real part of $1/a$,
in which case the imaginary part can be ignored except in the numerator
where it appears as a multiplicative factor.
lnelastic atom-atom scattering channels are also decay channels for the dimer.
The universal expression for the decay rate is
\be
\Gamma_2 = \frac{4 \hbar \, \mathrm{Im}[1/a]}{m \, a} .
\label{eq:Gamma2}
\ee
The  imaginary part of $a$ should be ignored  in the denominator.
The energy $\hbar \Gamma_2$ is twice the imaginary part of the complex binding energy given by Eq.~\eqref{eq:E2}
with complex $a$.  
Note that the imaginary part of $1/a$ cancels in the ratio of the inelastic cross section in Eq.~\eqref{eq:sigmain}
and the decay rate in Eq.~\eqref{eq:Gamma2}.

\subsection{Dark matter parameters}\label{sec:param}

\begin{figure}[t]
\centering
\includegraphics[width=0.8\linewidth]{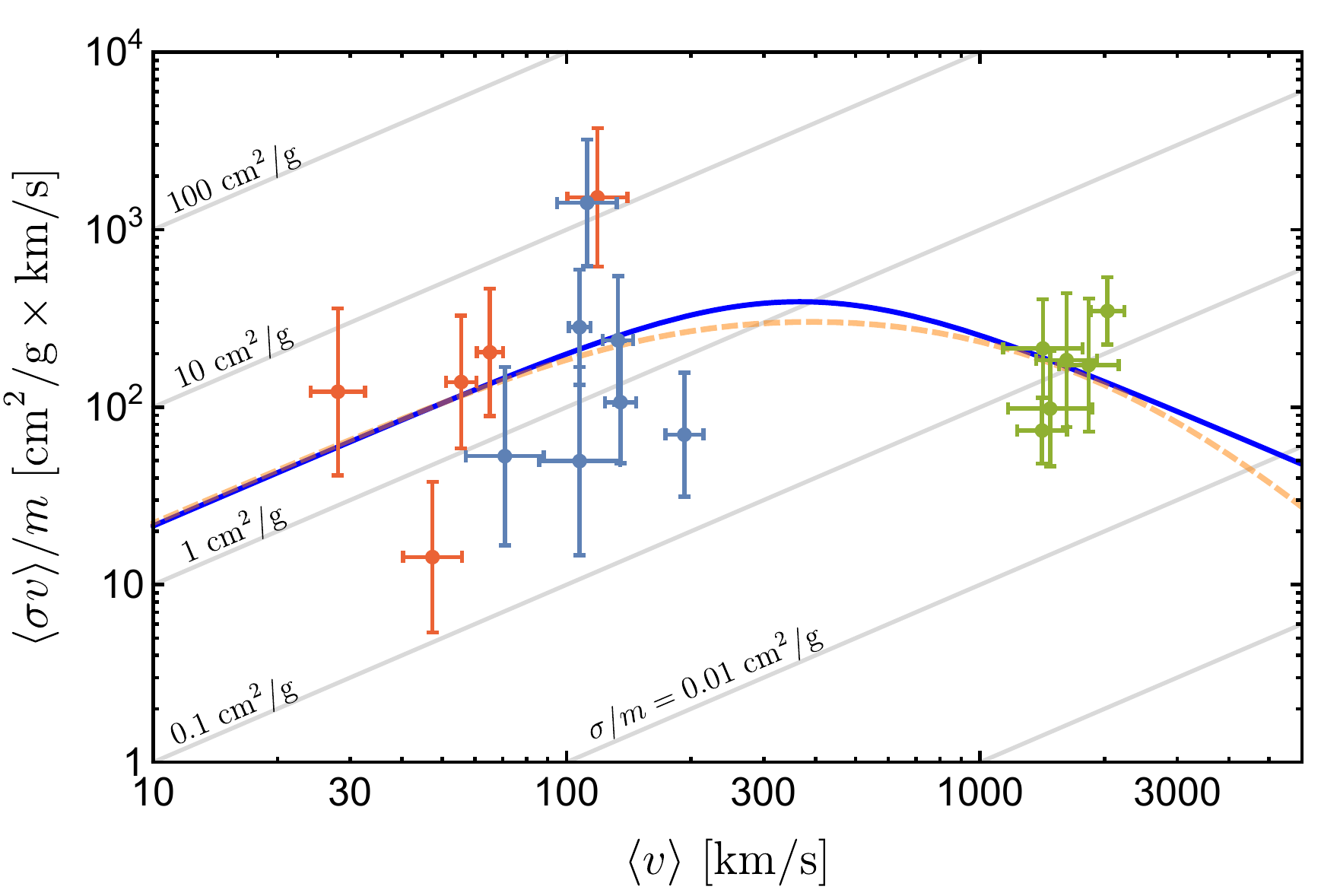}
\caption{Self-interaction reaction rate $\langle v\, \sigma_\mathrm{elastic}\rangle$ for dark matter particles 
as a function of  the  mean velocity $\langle v\rangle$.
The data points are results from Kaplinghat, Tulin, and Yu for dwarf galaxies (red), low-surface-brightness galaxies (blue), 
and galaxy clusters (green) \cite{Kaplinghat:2015aga}.
The curves are the best fit for a dark-photon model with $\alpha'=1/137$  \cite{Kaplinghat:2015aga}
(dashed) and the best fit to Eq.~\eqref{eq:vsigma} (solid).
The diagonal lines are for energy-independent cross sections.  
}
\label{fig:vsigma-v}
\end{figure}

The small-scale structure problems of the universe can be solved by 
self-interacting dark matter that becomes strongly interacting at low energies \cite{Rocha:2012jg, Peter:2012jh, Tulin:2017ara, Elbert:2016dbb}.
In Ref.~\cite{Kaplinghat:2015aga}, Kaplinghat, Tulin, and Yu determined 
self-interaction reaction rates $\langle v\, \sigma_\mathrm{elastic}\rangle$ for dark matter particles from astrophysical data on
dwarf galaxies, low-surface-brightness galaxies, and galaxy clusters\,\cite{Newman:2012nw, Newman:2012nv, Oh:2010ea, KuziodeNaray:2007qi}.
Their data points are shown as a function of the  mean relative velocity $\langle v\rangle$ 
of the dark atoms in Figure~\ref{fig:vsigma-v}.
In the galaxies, $\langle v\rangle$ ranges from about 20~km/s to about 200~km/s.
The values of $\langle v\, \sigma_\mathrm{elastic}\rangle$ for the galaxies only are roughly compatible 
with an energy-independent cross section with $\sigma_\mathrm{elastic}/m =2~\mathrm{cm}^2$/g.
In the galaxy clusters,  $\langle v\rangle$ is about 2000~km/s.
The values of $\langle v\, \sigma_\mathrm{elastic}\rangle$ for the clusters only 
are compatible with an energy-independent cross section
with $\sigma_\mathrm{elastic}/m =0.1~\mathrm{cm}^2$/g.
To fit the results  for both the galaxies and the clusters requires a cross section
that increases dramatically with decreasing velocity.
The results for $\langle v\, \sigma_\mathrm{elastic}\rangle$  versus $\langle v\rangle$
can be fit by a dark-photon model with three parameters:  the dark matter mass $m_\chi$, 
the dark photon mass $\mu$, and the coupling constant $\alpha'$ for a Yukawa potential.  
Kaplinghat et al.\ included additional systematic errors
of  0.3 in $\log(\langle v\, \sigma_\mathrm{elastic}\rangle/m)$  and 0.1 in $\log(\langle v\rangle)$  for each system
to take into account the uncertainty in their modeling.
They fixed the coupling constant at $\alpha' = 1/137$
and fit the parameters $m_\chi$ and $\mu$.
Their fitted values are $m_\chi=15^{+7}_{-5}$~GeV and $\mu = 17\pm 4$~MeV.
The curve for their best  fit with $m_\chi =15$~GeV and $\mu = 17$~MeV is shown in Figure~\ref{fig:vsigma-v}.

The results for the self-interaction reaction rates in Ref.~\cite{Kaplinghat:2015aga}
can be fit equally well by a short-range interaction model with a large scattering length.  
We assume dark nucleons are identical spin-0 bosons 
with a large real and positive scattering length.
The parameters required to describe the universal two-body physics
of dark nucleons are their mass $m_\chi$ and the scattering length $a$.
The elastic cross section is given in Eq.~\eqref{eq:sigma}.
The reaction rate as a function of the relative velocity $v$ is 
\be
v\, \sigma_\mathrm{elastic}(v) =\frac{8 \pi a^2v}{1 + (a m_\chi/2)^2 v^2} .
\label{eq:vsigma}
\ee
Our fit to the data points for $\langle v\, \sigma_\mathrm{elastic}\rangle$  versus $\langle v\rangle$
shown in Figure~\ref{fig:vsigma-v} gives
\begin{subequations}
\begin{align}
m_\chi &=19^{+3}_{-2}~\text{GeV}\,, 
\label{eq:mass}
\\
a& = \pm (17 \pm 3)~\text{fm}\,.
\label{eq:a}
\end{align}
\label{eq:massa}%
\end{subequations}
The curve for the best fit with $m_\chi=19$~GeV and $a = \pm 17$~fm is shown in Figure~\ref{fig:vsigma-v}.
The binding energy $E_2=1/(m_\chi a^2)$ of the dark deuteron is predicted to be   7.1~keV.
This is also the value of the collision energy $m_\chi v^2/4$
where $v\, \sigma_\mathrm{elastic}(v)$ has a maximum as a function of $v$.
The relative velocity $v$ at the maximum is about 300~km/s.
The reaction rate in Eq.~\eqref{eq:vsigma} must remain accurate for $v$ beyond 
the values of $\langle v \rangle$ for galaxy clusters.
At some larger velocity scale $v_0 = 2/m_\chi r_0$ set by the range $r_0$ of self-interactions,
 the reaction rate in Eq.~\eqref{eq:vsigma} may cross over to that for an energy-independent cross section,
 which is a diagonal line in Figure~\ref{fig:vsigma-v}.  Assuming the crossover does not occur until $v$ is at least 
 3 times larger than $\langle v \rangle$ for galaxy clusters,
 the energy-independent cross section must satisfy $\sigma_\mathrm{elastic}/m_\chi < 0.01~\mathrm{cm}^2$/g.
The range $r_0$ must be less than 0.5~fm,
and the energy scale $E_0 = 1/m_\chi r_0^2$ set by the range must be greater than 200~MeV.

One can obtain a very similar curve in Fig. 1 with spin-$\tfrac12$ fermions, for which the
factor $8\pi$ in \eq{vsigma} is replaced by  $4\pi$. The best-fit parameters for the mass and scattering length are 15~GeV and $\pm 22$~fm. This mass is the same as that obtained in the dark photon model of Ref.~\cite{Kaplinghat:2015aga}. The same scattering length could be obtained in that model by tuning either the dark photon mass or the Yukawa coupling constant. The mapping from the parameters of a model of dark fermions with gauge bosons to the scattering length is extensively discussed in Refs.~\cite{Braaten:2017gpq,Braaten:2017kci,Braaten:2017dwq}. The mapping from the parameters of a more fundamental dark matter model with bosons to the scattering length could be as nontrivial as the mapping from the parameters of QCD to the large neutron scattering length.

An upper bound on the elastic cross section for dark matter particles
has been obtained from the Bullet Cluster, which is the result of a collision of two galaxies 
with a relative velocity  estimated to be $\mathcal{O}$(1000)~km/s\,\cite{Clowe:2006eq, Springel:2007tu, Lee:2010hja, Kahlhoefer:2013dca, Lage:2013yxa, Lage:2014yxa, Kraljic:2014soa}.
The apparent absence of significant scattering from the two dark matter halos 
implies an upper bound on the elastic cross section for the dark matter particles
divided by their mass. 
If the dark matter particles have an energy-independent cross section, the upper bound 
on $\sigma_\mathrm{elastic}/m_\chi$  is  roughly $1~\mathrm{cm}^2$/g\,\cite{Massey:2017cwf, Harvey:2015hha, Robertson:2016xjh}.
The curves in Figure~\ref{fig:vsigma-v} at $\langle v \rangle = 1000$~km/s
are compatible with this bound.

Another constraint on the elastic cross section for dark matter particles
can be obtained by demanding that self-scattering removes 
the cusp in the dark matter distribution at the center of dwarf galaxies 
that is predicted by the $\Lambda$CDM model. 
If the dark matter particles have an energy-independent cross section, this condition provides 
an estimate of $\sigma_\mathrm{elastic}/m_\chi$ that is roughly $1~\mathrm{cm}^2$/g\,\cite{Tulin:2017ara}.
A typical mean velocity of dark matter particles in a dwarf galaxy is 10~km/s.
The curves in Figure~\ref{fig:vsigma-v} at $\langle v \rangle = 10$~km/s are compatible with this estimate.

Note that in our fit to \eq{vsigma}  we have only considered the data compiled by Kaplinghat et al.\ 
in Ref.~\cite{Kaplinghat:2015aga}.  We are aware that there are analyses of astrophysical systems 
that exhibit agreement with collisionless dark matter (see e.g.\  Refs.~\cite{Massey:2017cwf,Read:2018pft}).  There are caveats to these analyses, since the galaxy-dark matter offsets are 
predicted by strongly interacting dark matter are small \cite{Kim:2016ujt} and since the concentration 
parameter of dwarf galaxies may be higher than assumed\footnote{private communication from M.~Kaplinghat}.
Whether or not self-interacting dark matter is required in astrophysical systems requires more research. 
We will use the fit parameters in \eq{massa} to illustrate the near-threshold S-wave resonance model, 
but we will also consider values of the parameters that do not solve the small-scale structure problems.

\section{Universal three-body physics of identical bosons}
\label{sec:three-body}

In this section, we summarize the universal three-body physics of identical bosons with a large scattering length,
which is surprisingly intricate \cite{Braaten:2004rn}.
The 3-body physics depends strongly on the scattering length $a$.
It also depends log-periodically on a 3-body parameter $\kappa_*$ 
that can be determined from the binding energy of a universal bound 3-body cluster.

 In this section, 
we  use the concise  language of atomic physics for the particles and their bound clusters.  
The particle $d$ is referred to as an {\it atom}.
A  two-body bound cluster $d_2$, a three-body  bound cluster $d_3$, and a  four-body bound cluster $d_4$
are called a {\it dimer}, a {\it trimer}, and a {\it tetramer}, respectively.
We make factors of Planck's constant $\hbar$ explicit.

\subsection{Trimer spectrum}

The remarkable nature of trimers composed of identical bosons 
with a large scattering length was first realized by Vitaly Efimov.
In 1970, Efimov pointed out that in the unitary limit where $a$ is infinitely large,
there is a sequence of infinitely many trimers whose binding energies have  an accumulation point
at the 3-atom scattering threshold \cite{efimov1970energy}.
The ratio of the binding energies of two successive trimers is the square of 
a universal number $\lambda_0 = 22.694$.
The order of magnitude of the binding energy of the most deeply bound 
Efimov trimer is the energy scale 
$E_0 = \hbar^2/m r_0^2$ set by the range.

The discrete spectrum of Efimov trimers in the unitary limit $a = \pm \infty$
implies that few-body physics in the zero-range limit 
must depend on a 3-body parameter.
The Efimov trimers can be labeled by an integer $n$.
A convenient choice for the 3-body parameter is the binding  wave number
$\kappa_*$ in the unitary limit $a = \pm \infty$ 
of some arbitrarily chosen Efimov trimer labelled by $n_*$.
In the unitary limit, the binding energies of the other  Efimov trimers
differ by integer powers of  $\lambda_0^2 \approx 515$:
\be
E_{3,n} = \lambda_0^{-2(n-n_*)} \hbar^2 \kappa_*^2 /m~~~{\rm at}~~~a = \pm \infty.
\label{eq:E3unitarity}
\ee
If the binding wave number of a different Efimov trimer was
chosen as the 3-body parameter, the value of $\kappa_*$
would differ by a multiplicative factor that is an integer power of $\lambda_0$.
Since $\kappa_*$ can only be defined modulo multiplicative factors of 
$\lambda_0$, few-body physics can only depend log-periodically on 
$\kappa_*$.  In particular, 3-body reaction rates must be functions of $a$ and 
$\kappa_*$ that are invariant under replacing $\kappa_*$ 
by $\lambda_0 \kappa_*$.

The binding energies of  Efimov trimers are smooth functions 
of the inverse scattering length $1/a$ \cite{efimov1973energy}.  
If the scattering length is not infinitely large, 
there are only a finite number of Efimov trimers.
As $1/a$ decreases through negative values, Efimov trimers disappear 
through the 3-atom scattering threshold at critical values of $a$
that differ by multiplicative factors of $\lambda_0$.
As $1/a$ increases through positive values, Efimov trimers disappear 
through the atom-dimer scattering threshold at critical values of $a$
that differ by multiplicative factors of $\lambda_0$.
The Efimov trimer whose binding momentum in the unitary limit 
is $\kappa_*$ disappears through the 3-atom scattering threshold 
$E=0$ at the negative scattering length  $a_- = -1.508~\kappa_*^{-1}$   \cite{2008PhRvL.100n0404G},
and it disappears through the atom-dimer scattering threshold $E=-E_2$ 
at the positive scattering length $a_* \approx 0.07076~\kappa_*^{-1}$ \cite{Braaten:2004rn}.
Given any large scattering length $a$,
the energy of one Efimov trimer can be used to determine $\kappa_*$
and the binding energies of the other Efimov trimers can then be predicted.
The number of Efimov trimers is not predicted, 
because the binding energy of the deepest Efimov trimer is 
determined by the range $r_0$.

In atomic physics, the two $^4$He trimers are beautiful examples of Efimov trimers.   
The binding energy of the more deeply bound 
$^4$He trimer is about $1.1\times 10^{-5}$~eV.
It was first observed using diffraction from a transmission grating \cite{1994Sci...266.1345S}.
The binding energy of the more weakly bound $^4$He trimer 
relative to the 3-atom threshold is  about $2.3 \times 10^{-7}$~eV,
which is about a factor of 2 larger than that of the $^4$He dimer. 
It has been observed only recently
using Coulomb explosion imaging \cite{voigtsberger2014imaging}.
The first Efimov trimer observed in cold atom physics was a $^{133}$Cs trimer observed in 2008
as a resonance in the atom loss rate from 3-body recombination \cite{kraemer2006evidence}. 

\subsection{Dimer-atom scattering}

The dimer-atom scattering processes are elastic scattering
($d_2+d \to d_2+d$) and dimer-breakup scattering ($d_2+d \to d+d+d$).
The collision energy, which is the total kinetic energy of the atom and dimer in the center-of-momentum frame, is 
\be
E =  \frac{3 \hbar^2 k^2}{4m},
\label{eq:Ecm}
\ee
where $\hbar k$ is the relative momentum of the atom and dimer.
The partial wave expansion for the elastic scattering amplitude is
\be
f_k(\theta) = \sum_{J=0}^\infty
\frac{2J+1}{k \cot \delta_J(k) - i k} P_J(\cos \theta).
\label{eq:fktheta}
\ee
The phase shifts $\delta_J(k)$ are dimensionless functions of $k$.
The scattering is purely elastic for energies between the atom-dimer threshold 
$E=0$ and the dimer-breakup threshold $E=E_2$.
The phase shifts are therefore  real for $E<E_2$ and complex for $E>E_2$.
The cross sections for elastic scattering and for breakup scattering
can be expressed in terms of the phase shifts:
\begin{subequations}
\begin{align}
\sigma_{\rm elastic}(E) &= 
\frac{4 \pi}{k^2} \sum_{J=0}^\infty (2J+1)
\left| e^{i \delta_J(k)} \sin \delta_J(k) \right|^2\,, 
\label{eq:sigmaelastic}
\\
\sigma_{\rm breakup}(E) &= 
\frac{ \pi}{k^2} \sum_{J=0}^\infty (2J+1)
\left( 1 - \big| e^{2 i \delta_J(k)} \big|^2 \right)\,.
\label{eq:sigmabreakup}
\end{align}
\label{eq:sigmaAD}
\end{subequations}

\begin{figure}[t]
\centering
\includegraphics[width=0.8\linewidth]{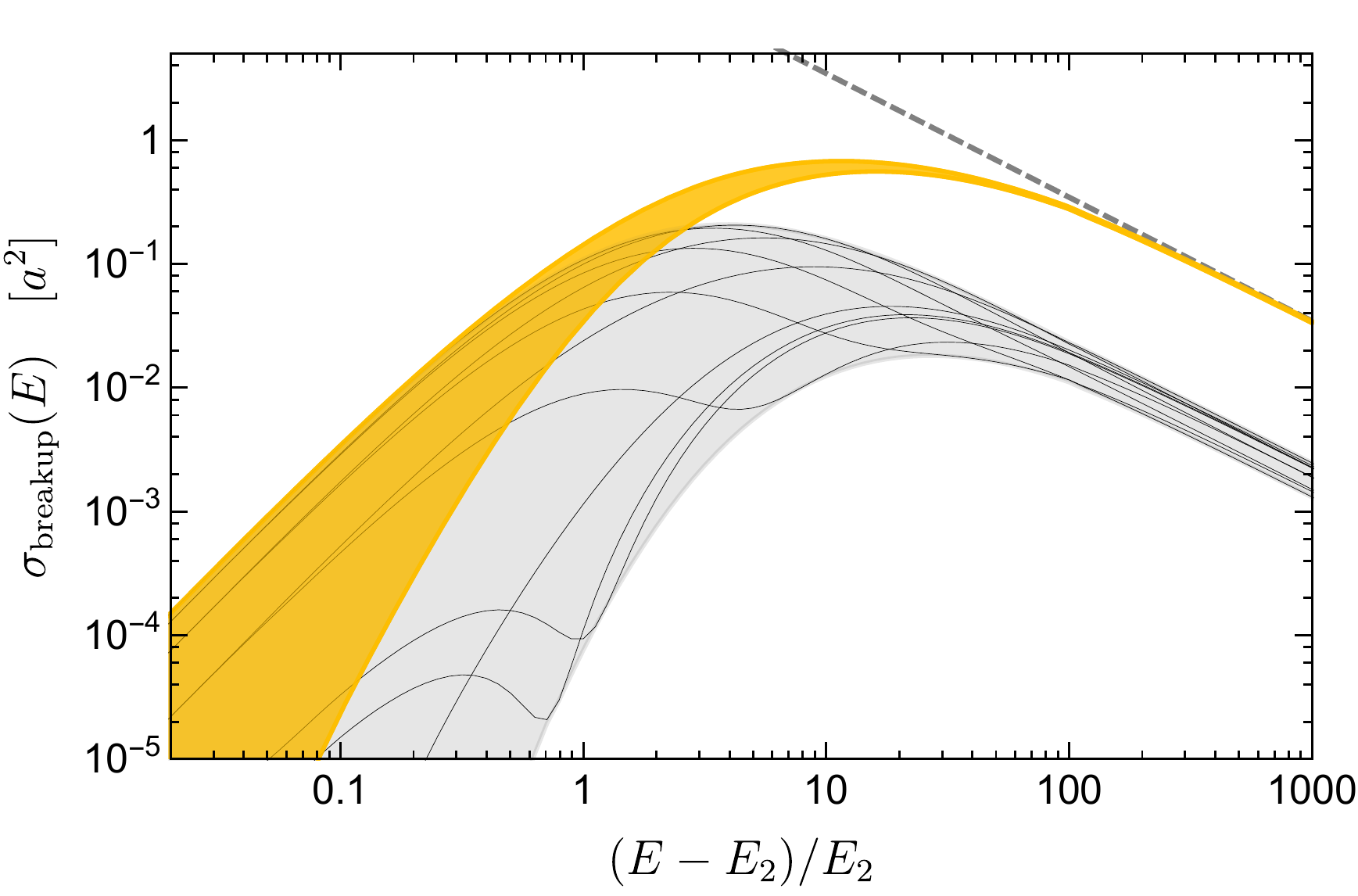}
\caption{
The breakup cross section as a function of energy with respect to the threshold: $E-E_2$.
The upper band is the  envelope of $\sigma_{\rm breakup}(E)$ for all possible values of  the three-body parameter $a_+$.
The lower  band is the envelope of the $J=0$ contribution to $\sigma_{\rm breakup}(E)$ for all possible  values of $a_+$. 
The curves inside the lower band are for 
8 values of $a_+$: $a_+/a= \lambda_0^{n/8}$, $n=0,1,\ldots,7$.
The dashed line is the extrapolation from the scaling behavior in \eq{sigscaling}.
}
\label{fig:sigbrk}
\end{figure}

In the universal regime where the energy $E$ is much smaller than 
the energy scale $E_0$ set by the range,
the only relevant interaction parameters are the scattering length $a$ 
and the 3-body parameter $\kappa_*$.
The S-wave phase shift $\delta_0(k)$ is a dimensionless function of 
$ka$ and $a \kappa_*$ 
that depends only log-periodically on $a \kappa_*$.
It can be expressed in the form  \cite{Braaten:2004rn}
\be
\exp \big( 2 i \delta_0(E)\big) = s_{22}(x)+ 
\frac{ s_{12}(x)^2 \exp[2i s_0 \log(a/a_+)]}
{1 - s_{11}(x) \exp[2i s_0 \log(a/a_+)]} ,
\label{eq:sigmaelbreakup}
\ee
where $s_0 = \pi/\log\lambda_0 \approx 1.00624$ is a universal constant
and $a_+$ is an alternative 3-body parameter that differs from $\kappa_*^{-1}$ by a multiplicative factor:
$a_+ = 0.3165\,\kappa_*^{-1}$ \cite{2008PhRvL.100n0404G}.
The dimensionless functions $s_{11}(x)$, $s_{12}(x)$, and $s_{22}(x)$ 
are complex-valued  functions of the scaling variable $x = ka$. 
In terms of this variable, the dimer-breakup threshold $E=E_2$ is $x=2/\sqrt{3}$.
For $x < 2/\sqrt{3}$, the scaling functions $s_{11}(x)$, $s_{12}(x)$, and $s_{22}(x)$
are entries of a $2\times2$ unitary matrix.
For $x > 2/\sqrt{3}$, they are entries of a $2\times 2$ submatrix of a $3\times3$ unitary matrix.
In Ref.~\cite{Braaten:2008kx}, they were calculated numerically
over the range of $x$ from $10^{-1}$ to $10^{+1}$.
The phase shifts $\delta_J(k)$ for the higher partial waves are 
also dimensionless functions of $x =ka$.
They are real for $x < 2/\sqrt{3}$ and complex for $x > 2/\sqrt{3}$.
In Ref.~\cite{Braaten:2008kx}, the phase shifts for 
$J$ from 1 to 6 were calculated numerically
over the range of $x$ from $10^{-1}$ to $10^{+1}$.
The results of Ref.~\cite{Braaten:2008kx}
allow the elastic cross section and the breakup cross section 
to be calculated for collision energies up to
100~$E_2$, where $E_2$ is the dimer binding energy in Eq.~\eqref{eq:E2}.

The breakup cross section is shown in Figure~\ref{fig:sigbrk} as bands whose  
envelope corresponds to minimizing and maximizing the cross section with respect to $a_+$.
The upper band is the total cross section, and the lower band is the contribution from $J=0$.
The curves inside the lower band are for 8 values of $a_+/a$ between 1 and $\lambda_0$
that are equally spaced on a log scale.
At $E=100\,E_2$, the sum of the higher partial waves is larger than the maximum $J=0$ contribution 
by more than an order of magnitude. 
The behavior of the individual partial waves at large $E$ is consistent with decreasing as $1/E^2$,
but the sum of the partial waves is consistent with decreasing as $1/E$.
This is the scaling behavior of the cross section at high energy that is expected from dimensional analysis,
given that the only energy scale from interactions is $E_2$.
Both the upper and lower bands are extrapolated beyond 100~$E_2$ by fitting to power-law scalings 
between 50~$E_2$ and 100~$E_2$.  
The breakup cross section in the high-energy limit can be approximated as
\be
\sigma_\text{breakup}(E)\to c_2 \frac{\hbar^2}{mE}
\,,\label{eq:sigscaling}
\ee
where the coefficient is $c_2 \approx 35$.

Analytic expressions for the cross sections are known 
at special values of the energy \cite{Braaten:2004rn}.
They can be useful for making order-of-magnitude estimates.
The elastic cross section at the atom-dimer threshold is
\be
\sigma_{\rm elastic}(E = 0) = 
4 \pi \big( 1.46 + 2.15 \cot[s_0 \log(a/a_*)] \big)^2 a^2,
\label{eq:sigmathreshold}
\ee
where $a_* = 0.0708\, \kappa_*^{-1}$.
The cross section at $E=0$ depends log-periodically on $a/a_*$, and  its value ranges  from 0 to $\infty$.
It diverges at $a_* = a$ and at other values of $a_*$ 
that differ from $a$ by an integer power of $\lambda_0=22.7$, 
because there is an Efimov trimer at the atom-dimer threshold.
The cross section at $E=0$ vanishes at $a_* = 2.63\, a$ and at other values of $a_*$ 
that differ from $2.63\,a$ by an integer power of $\lambda_0$, 
because there is destructive interference between two scattering pathways.
The S-wave contribution to the elastic cross section at the dimer-breakup threshold is
known analytically.  It can be approximated with an accuracy of better than 1\% by \cite{Braaten:2004rn}
\be
\sigma_{\rm elastic}^{(J=0)}(E= E_2)  \approx
3\pi\, \sin^2[s_0 \log(a/a_+)] \, a^2,
\label{eq:sigmael0}
\ee
where $a_+ \approx 0.3165\,\kappa_*^{-1}$.
 This contribution vanishes at  $a_+ = a$ and at other values of $a_+$ that differ from $a$ 
 by an integer power of $\lambda_0$, because there is perfect destructive interference  between two reaction pathways.
The S-wave contribution to the breakup cross section 
at an energy $E$ just above the dimer-breakup threshold $E_2$ is \cite{Braaten:2008kx}
\be
\sigma^{(J=0)}_{\rm breakup}(E)\approx
\frac{C_3(a/a_+)}{32 \sqrt{3}\, \pi}\,
\left( \frac{E-E_2}{E_2} \right)^2 a^2.
\label{eq:sigmabreakup0}
\ee
The coefficient $C_3(a/a_+)$ in the prefactor depends log-periodically on $a$.
A completely analytic expression for this coefficient has been derived by 
Macek, Ovchinnikov, and Gasaneo \cite{macek2006exact}.
It can be expressed as
\be
C_3(a/a_+) = 
\frac{66.6373\, \sin^2[s_0 \log(a/a_+)]}{1- 0.00717\,\sin^2[s_0 \log(a/a_+)]}\,,
\label{eq:C3}
\ee
where $s_0 = \pi/\log\lambda_0 = 1.00624$ 
and $a_+ \approx 0.3165\,\kappa_*^{-1}$  \cite{2008PhRvL.100n0404G}.
It vanishes at $a_+=a$ and at other values of $a_+$ that differ from $a$ by an integer power of $\lambda_0=22.7$.
At these values of $a_+$, there is perfect destructive interference between two recombination pathways.
The coefficient in \eqref{eq:C3} can be approximated with an error of less than 1\% by the simpler expression
\begin{equation}
C_3(a/a_+) \approx 67.1 \sin^2[s_0 \log(a/a_+)].
\label{eq:C3approx}
\end{equation}

\subsection{Three-body recombination}

Three-body recombination is a reaction in which the collision of three atoms 
results in the formation of a dimer: $d+d+d \to d_2+d$.
The reaction rate depends on the wave vectors $\bm{k}_1$, $\bm{k}_2$, and $\bm{k}_3$
of the three colliding atoms, but not on the total wave vector $\bm{k}_1+\bm{k}_2+\bm{k}_3$.  
It can be expressed as a function of the Jacobi wave vectors defined 
by $\bm{k}_{12} = \bm{k}_1-\bm{k}_2$ and  $\bm{k}_{3,12} = \bm{k}_3-\tfrac12(\bm{k}_1+\bm{k}_2)$.
The collision energy $E$ is the kinetic energy in the center-of-momentum frame:
\be
E = \frac{\hbar^2 (3k_{12}^2+4k_{3,12}^2)}{12m}.
\label{eq:Ecm3}
\ee
The recombination rate can be expressed as a  function of $E$ and 5 dimensionless hyperangles 
consisting of the spherical angles of the two Jacobi vectors and $\arctan(\sqrt{3}\,k_{12}/2k_{3,12})$.
The hyperangular average of the recombination rate is a function of $E$ only.  It can be expressed 
in terms of the breakup cross section at the kinetic energy $E_2+E$ \cite{Braaten:2008kx}:
\be
\big \langle R(\bm{k}_{12},\bm{k}_{3,12}) \big\rangle = 
\frac{192\sqrt{3}\, \pi \hbar^3(E_2+E)}{m^2 E^2} \,\sigma_{\rm breakup}(E_2+E).
\label{eq:<R>T}
\ee

In the universal regime where  the collision energy is much smaller than 
the energy scale $\hbar^2/m r_0^2$ set by the range,
the only relevant interaction parameters are the scattering length $a$ 
and the 3-body parameter $\kappa_*$.
A completely analytic expression for the three-body recombination rate 
at zero collision energy has been derived by 
Macek, Ovchinnikov, and Gasaneo \cite{macek2006exact}.
It can be expressed as
\be
R(E=0) = 6\,C_3(a/a_+)\, \hbar a^4/m.
\label{eq:RE=0}
\ee
The coefficient $C_3(a/a_+)$ depends log-periodically on $a/a_+$, 
and it can be approximated by the expression in Eq.~\eqref{eq:C3approx}.
The recombination rate at $E=0$ vanishes at $a_+=a$ and at other values of $a_+$ that differ from $a$ 
by an integer power of $\lambda_0$.
It has its maximum value 67.1~$\hbar a^4/m$
at $a_+=4.76\,a$ and at other values of $a_+$ that differ from $4.76\,a$ by an integer power of $\lambda_0$.

\subsection{Four-body physics and beyond}

In 2004, Platter, Hammer, and Meissner predicted the existence of 
universal 4-body bound clusters composed of identical bosons with large scattering length 
 \cite{Platter:2004he,Hammer:2006ct}.
Their binding energies were mapped out as functions of $a$ by 
von Stecher, D'Incao, and Greene 
\cite{2008arXiv0810.3876V}. 
In an experiment with $^{133}$Cs atoms in 2009, the dramatic increase of the 
4-body recombination rate at low temperature near a specific value of $a$
was used to discover the first such universal tetramer 
\cite{2009PhRvL.102n0401F}. 
There is theoretical evidence for universal bound clusters of 5, 6, and even more identical bosons
with a large scattering length \cite{2011PhRvL.107t0402V}.

\section{Rate Coefficients at Thermal Equilibrium}
\label{sec:thermal}

In this section, we give expressions for the rate coefficients for few-body reactions for identical bosons 
with large scattering lengths in thermal equilibrium.
We  use the concise language of atomic physics for the bosons and their bound clusters.
We consider a gas of atoms  with number density $n_1$
and dimers with number density $n_2$
in kinetic  equilibrium at temperature $T$
but not necessarily in chemical equilibrium.
For simplicity, we assume the gas is sufficiently dilute 
that the Bose-Einstein momentum distributions of the atoms and dimers
can be approximated by Maxwell-Boltzmann distributions.

\subsection{Inelastic atom-atom scattering}

\begin{figure}[t]
\centering
\includegraphics[width=0.8\linewidth]{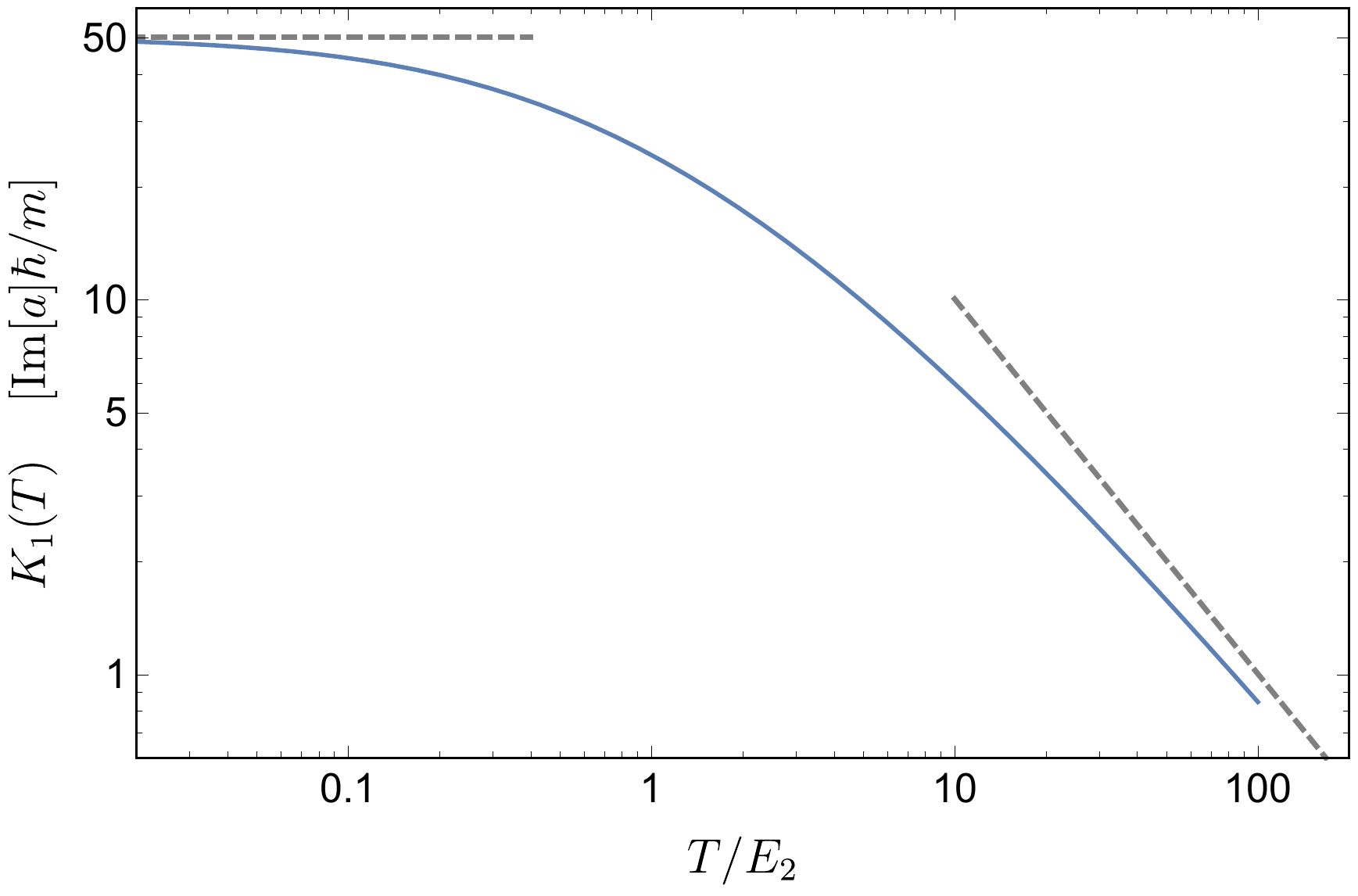}
\caption{
Rate coefficient $K_1(T)$ for dimer breakup as a function of the temperature $T$.
The dashed lines indicate its asymptotic behavior in the low-$T$ and high-$T$ limits.
}
\label{fig:K1}
\end{figure}

We assume all the inelastic atom-atom scattering channels have energy release large compared to $E_2$,
and that the energetic particles produced by the reaction have scattering cross sections 
with an atom that are small compared to the elastic atom-atom cross section.
The particles produced by an inelastic reaction can therefore be ignored,
and the only effect of the reaction is to decrease the number of atoms by 2.
In a homogeneous system, the rate at which the number density $n_1$ of atoms decreases
from inelastic atom-atom  scattering is proportional to $n_1^2$:
\be
\frac{d\ }{dt} n_1 = -2 K_1(T) \, n_1^2.
\label{eq:dn1/dt:K1}
\ee
The rate coefficient $K_1(T)$ depends on the temperature
and can be expressed as a weighted integral over the inelastic cross section:
\be
K_1(T) 
= \frac{4}{\sqrt{ \pi m}\, (kT)^{3/2}} 
\int _0^\infty dE\, E\, e^{-E/kT} \, \sigma_\mathrm{inelastic}(E).
\label{eq:K1T}
\ee
Upon inserting the universal approximation to the  inelastic cross section 
for identical bosons in Eq.~\eqref{eq:sigmain},
we obtain an analytic result:
\be
K_1(T) 
= \left(32\pi\, g(kT/E_2)\frac{E_2}{kT} \right)
\frac{ \hbar\, \mathrm{Im}[a]}{m}\,,
\label{eq:K1Tanalytic}
\ee
where the dimensionless function $g(t)$ is  $1-(\pi/t)^{1/2}e^{1/t} [1- \mathrm{erf}(1/\sqrt{t})]$.
Figure~\ref{fig:K1} shows the rate coefficient $K_1(T)$ and its  limiting behaviors: 
$16\pi\,  \hbar\, \mathrm{Im}[a]/m$ in the low-$T$ limit and $32\pi (E_2/kT )\hbar\, \mathrm{Im}[a]/m$ in the high-$T$ limit.

\subsection{Dimer breakup}

The dimer-breakup reaction $d d_2 \to ddd$ decreases the number of dimers by 1 
and increases the number of atoms by 2.
We assume the final-state atoms are thermalized by the elastic atom-atom scattering.
In a homogeneous system,
the rate at which the number density $n_2$ of dimers decreases
from dimer-breakup scattering is proportional to $n_1 n_2$:
\be
\frac{d\ }{dt} n_2 = - K_2(T) \, n_1 n_2.
\label{eq:dn2/dt:K2}
\ee
The rate coefficient $K_2(T)$  depends on the temperature
and can be expressed as a Boltzmann average of  the dimer-breakup cross section:
\be
K_2(T) = \frac{6}{\sqrt{3 \pi m}\, (kT)^{3/2}}
\int _{E_2}^\infty dE\, E\,e^{-E/kT} \sigma_{\rm breakup}(E).
\label{eq:K2T}
\ee

\begin{figure}[t]
\centering
\includegraphics[width=0.8\linewidth]{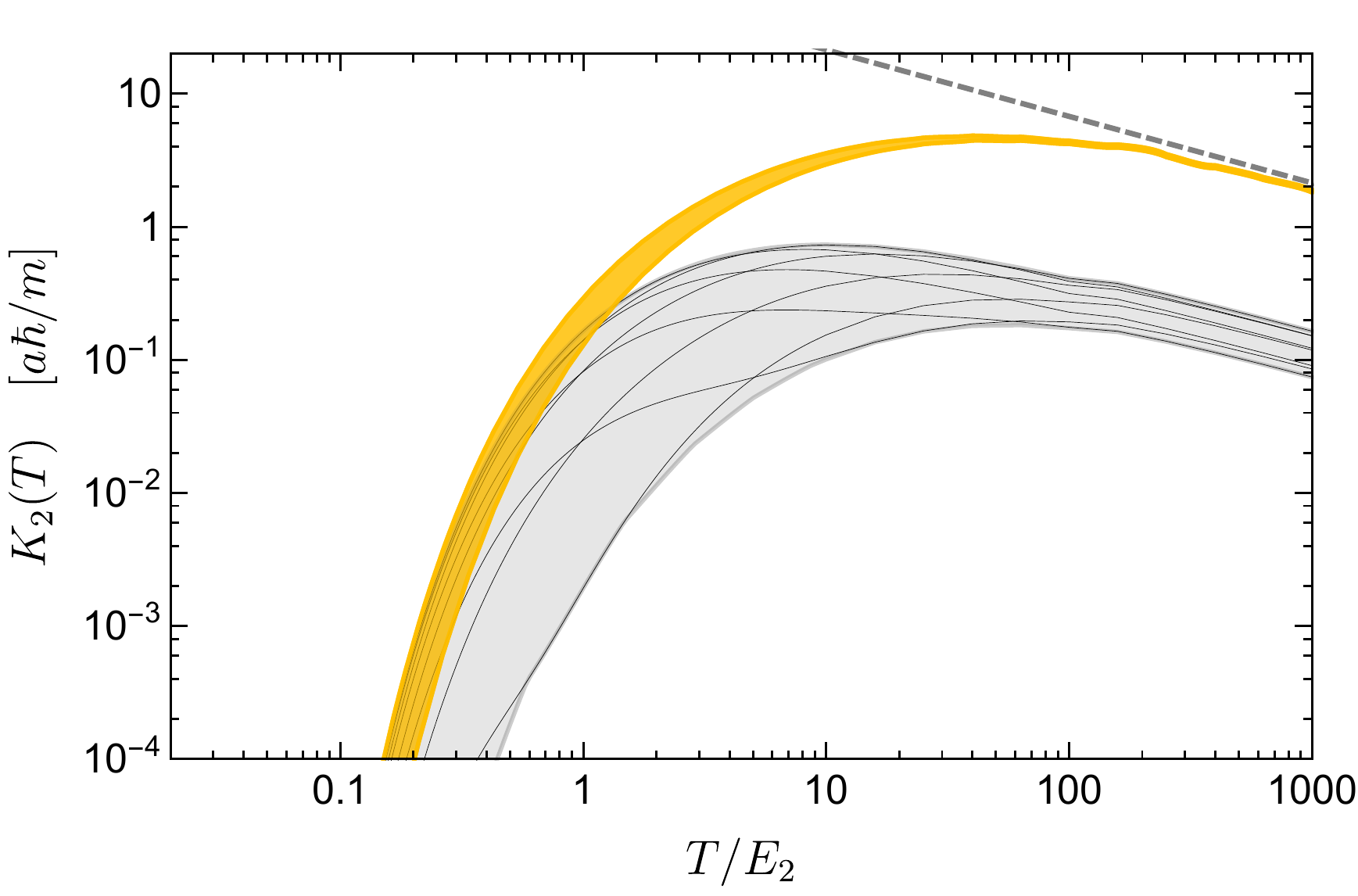}
\caption{
Rate coefficient $K_2(T)$ for dimer breakup as a function of the temperature $T$.
The upper band is the envelope of $K_2(T)$ for all possible values of the three-body parameter $a_+$. 
The dashed line is the extrapolation from the scaling behavior in \eq{K2scaling}.
The lower band is the envelope of the $J=0$ contribution to $K_2(T)$ for all possible values of $a_+$. 
The curves inside the lower band are for 8 values of $a_+$: $a_+/a= \lambda_0^{n/8}$, $n=0,1,\ldots,7$.}
\label{fig:K2}
\end{figure}

The universal approximation to the dimer-breakup cross section is given  in Eq.~\eqref{eq:sigmaAD}.
The universal results for the atom-dimer phase shifts $\delta_J(k)$ in Ref.~\cite{Braaten:2008kx}
are obtained up to about $100\; E_2$.  The breakup cross section 
is extended above $100\; E_2$ as shown in Figure~\ref{fig:sigbrk} by fitting to the power-law behavior
in Eq.~\eqref{eq:sigscaling}.
This allows $K_2(T)$ to be calculated for all temperatures $kT$ up to the scale $E_0$ set by the range.
The results for the rate coefficient are shown in Figure~\ref{fig:K2} as bands whose  
envelope corresponds to minimizing and  maximizing the rate with respect to $a_+$.
The upper band is the total rate coefficient, and the lower band is the contribution from $J=0$.
The curves inside the lower band are for 8 values of $a_+/a$ between 1 and $\lambda_0$
that are equally spaced on a log scale.

We can obtain a simple analytic approximation for $K_2(T)$ in the 
{\it low-temperature limit} $kT \ll E_2$. In this limit, 
the  breakup cross section  in \eq{sigmabreakup0} is dominated by the S-wave contribution.
The limiting behavior of the rate coefficient is
\begin{equation}
K_2(T) \longrightarrow 
\frac{C_3(a/a_+)}{2\sqrt{2}}\, e^{-E_2/kT}\left( \frac{a}{\lambda_T} \right)^3  \frac{\hbar a}{m}\,,\
\label{eq:K2threshold}
\end{equation}
where $\lambda_T=(2\pi \hbar^2/m kT)^{1/2}$ is the thermal wavelength
and $C_3(a/a_+)$ is the coefficient 
of $\hbar a^4/m$ in the 3-body recombination rate at zero collision energy in Eq.~\eqref{eq:RE=0}.
This coefficient can be accurately approximated by  Eq.~\eqref{eq:C3approx}.
Note that the dimer-breakup rate coefficient in Eq.~\eqref{eq:K2threshold} is exponentially suppressed by the Boltzmann factor.

We can also obtain a simple analytic approximation for $K_2(T)$ in the 
{\it scaling region}, where $kT$ is much larger than $E_2$
and much smaller than  the energy scale $E_0= \hbar^2/mr_0^2$ set by the range.
In the scaling region $E_2 \ll kT \ll E_0$,
the  breakup cross section  in \eq{sigmabreakup0} is dominated by the higher partial-wave contributions.
Figure~\ref{fig:K2} shows that at $kT = 100\,E_2$, the sum of the higher partial waves is already 
more than an order of magnitude larger than the  maximum $J=0$ contribution.
The dependence on the S-wave scattering length $a$ can therefore be neglected.
Since the interactions provide no other length scales smaller than the range, the dependence of
the rate coefficient on $T$ can be determined 
up to a numerical coefficient by dimensional analysis:
\begin{equation}
K_2(T) \longrightarrow c_2\frac{ \sqrt{6}}{\pi}\,  \frac{\hbar \lambda_T}{m},
\label{eq:K2scaling}
\end{equation}
where $c_2\approx 35$ is the same coefficient as in Eq.~\eqref{eq:sigscaling}.
The extrapolation in  $T$ provided by the scaling behavior in Eq.~\eqref{eq:K2scaling}
is shown as a dashed line in Figure~\ref{fig:K2}.

\subsection{Three-body recombination}

\begin{figure}[t]
\centering
\includegraphics[width=0.8\linewidth]{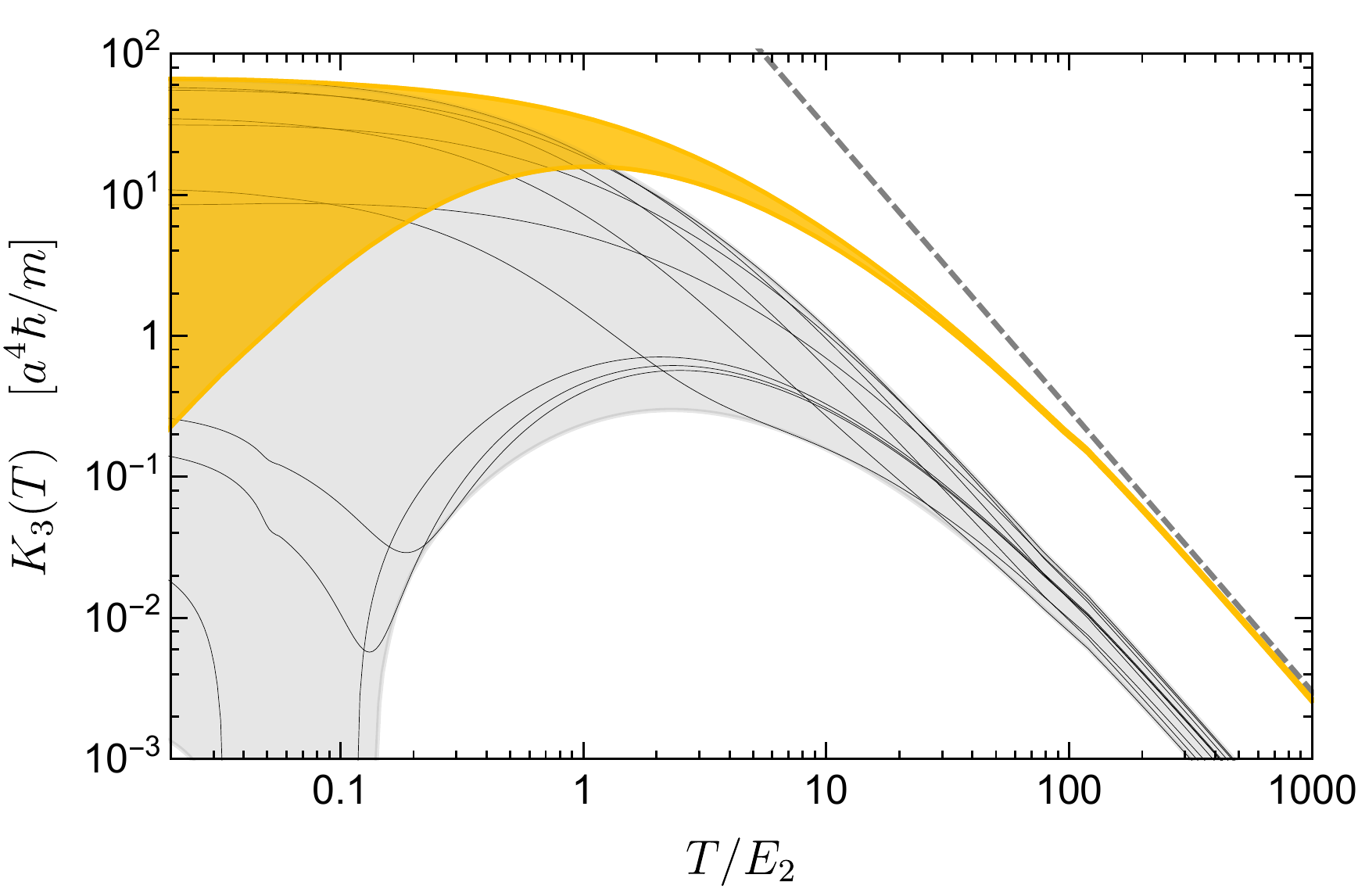}
\caption{
Rate coefficient $K_3(T)$ for three-body recombination as a function of the temperature $T$.
The upper band is the  envelope of $K_3(T)$ for all possible values of  the three-body parameter $a_+$.
The dashed line is the extrapolation from the scaling behavior in \eq{K3scaling}.
The lower  band is the envelope of the $J=0$ contribution to $K_3(T)$ for all possible  values of $a_+$. 
The curves inside the lower band are for 8 values of $a_+$: $a_+/a= \lambda_0^{n/8}$, $n=0,1,\ldots,7$.}
\label{fig:K3}
\end{figure}

The three-body recombination reaction $ddd  \to d d_2$ increases the number of dimers by 1 
and decreases the number of atoms by 2.
We assume the final-state atom and the final-state dimer are thermalized by elastic atom-atom scattering
and by elastic atom-dimer scattering, respectively.
In a homogeneous system,
the rate at which the number density $n_2$ of dimers increases
from 3-body recombination is proportional to $n_1^3$:
\be
\frac{d\ }{dt} n_2 = + K_3(T) \, n_1^3.
\label{eq:dn2/dt:K3}
\ee
The rate coefficient $K_3(T)$ depends on the temperature
and can be expressed as a Boltzmann average of  the three-body recombination rate:
\be
K_3(T) = \frac{\int _0^\infty dE\, E^2 \, e^{-E/kT}  \,  \big \langle R(\bm{k}_{12},\bm{k}_{3,12}) \big\rangle}
{6 \int _0^\infty dE\, E^2\,  e^{-E/kT}}, 
\label{eq:K3T-B}
\ee
where $\big \langle R\big\rangle$ is the hyperangular average of the 
3-body recombination rate, which is a function of the collision energy $E$ only.
The factor of $1/3!$ compensates for the overcounting of 3-body states of the 3 identical bosons
in the Boltzmann average.
The rate coefficient can be expressed as a weighted  integral over the dimer-breakup cross section:
\be
K_3(T) = \frac{16\sqrt{3}\, \pi \hbar^3}{m^2 (kT)^3}
\int _0^\infty dE\, e^{-E/kT}  \,  (E_2+E)\,\sigma_{\rm breakup}(E_2+E).
\label{eq:K3T}
\ee
We can use \eq{K2T} to express $K_3(T)$ in terms of $K_2(T)$:
\be
K_3(T) =  2\sqrt{2}\, \lambda_T^3 e^{E_2/kT}\, K_2(T)\,,
\label{eq:K3T-K2T}
\ee
where $\lambda_T=(2\pi \hbar^2/m kT)^{1/2}$ is the thermal wavelength.
The universal approximation to the dimer-breakup cross section is given  in Eq.~\eqref{eq:sigmaAD}.
The universal results for the atom-dimer phase shifts $\delta_J(k)$ in Ref.~\cite{Braaten:2008kx}
are obtained up to about $100\; E_2$ and the breakup cross section 
is extended above $100\; E_2$ in Figure~\ref{fig:sigbrk} by fitting to the power-law behavior.
This allows the recombination rate coefficient $K_3(T)$ to be calculated for all temperatures
$kT$ up to scale $E_0$ set by the range.
The results are shown in Figure~\ref{fig:K3} as bands whose 
envelopes corresponding to minimizing and  maximizing  the rates with respect to $a_+$.
The upper band is the total rate coefficient, and the lower band is the contribution from $J=0$.
The curves inside the lower band are for 8 values of $a_+/a$ between 1 and $\lambda_0$
that are equally spaced on a log scale.

We can obtain simple analytic approximations for the 3-body recombination rate coefficient
 by using the relation between $K_2(T)$ and $K_3(T)$ in  \eq{K3T-K2T}
and the analytic approximations for $K_2(T)$  in Eqs.~\eqref{eq:K2threshold} and \eqref{eq:K2scaling}.
In the  low-temperature limit $kT \ll E_2$,
the  rate coefficient approaches a constant that depends log-periodically on $a/a_+$:
\be
K_3(T) \longrightarrow C_3(a/a_+)\,  \frac{\hbar a^4}{m} \,,
\label{eq:K3threshold}
\ee
where $C_3(a/a_+)$ is the coefficient in Eq.~\eqref{eq:C3}, which
can be accurately approximated by Eq.~\eqref{eq:C3approx}.
In the scaling region $E_2 \ll kT \ll E_0$, 
the  rate coefficient scales as the power of temperature required by dimensional analysis:
\be
K_3(T) \longrightarrow c_2\frac{ 4\sqrt{3} }{\pi}\,  \frac{\hbar\lambda_T^4}{m} \,,
\label{eq:K3scaling}
\ee
where $c_2 \approx 35$ is the same coefficient as in Eq.~\eqref{eq:sigscaling}.
The extrapolation in  $T$ provided by the scaling behavior in Eq.~\eqref{eq:K3scaling}
is shown as a dashed line in Figure~\ref{fig:K3}.

In experiments with ultracold trapped atoms,
the atoms form an extremely dilute gas
in the sense that the typical interatom spacing is much larger than the 
range of the interactions between atoms:
$\langle n_1\rangle^{1/3} r_0 \ll 1$, where $\langle n_1\rangle$
is the density-weighted average of the number density.
Three-body recombination can be important in these experiments,
because the dimer and atom in the final state often have enough kinetic energy 
to escape from the trapping potential. 
In that case, every recombination event results in the loss of three atoms.
In an experiment with $^{133}$Cs atoms in 2005, the dramatic increase of the 
3-body recombination rate at low temperature 
when the scattering length was tuned to near the negative value $a_- = -1.5\,\kappa_*^{-1}$
was used to discover an Efimov trimer \cite{kraemer2006evidence}.


\section{Early universe}
\label{sec:universe}

In this section, we study the production of dark deuterons through three-body recombination 
of dark nucleons during the Hubble expansion of the early universe
under the assumption that the dark nucleons are identical bosons with a large positive scattering length.
We calculate the fraction of dark matter in the form of dark  deuterons as a function of the redshift.

\subsection{Rate equations}\label{sec:rate-eq}

After the decoupling of dark matter from ordinary matter, the densities 
of dark nucleons and larger dark nuclei evolve in thermal equilibrium until
they are captured by the gravitational potential wells of galaxies.
The time evolution is due to the Hubble expansion and to reactions among 
the dark nuclei.  Assuming that the larger dark nuclei are weakly bound,
the density and temperature at decoupling are large enough that any
larger dark nucleus that is formed is immediately broken up by a collision
with a dark nucleon.  Thus we can take as an initial condition 
that the dark matter consists entirely of dark nucleons at the decoupling time.

Given an initial state consisting only of dark nucleons,
larger dark nuclei can be formed by $N$-body recombination reactions
in which $N$ dark nuclei collide and some of them form bound states.
At sufficiently low dark nucleon number density $n_1$, 
the $N$-body recombination rate is proportional to $n_1^N$.
Thus if a dark deuteron $d_2$ exists, the  most favorable reaction is
3-body recombination ($d+d+d\leftrightarrow d_2 +d$).
Once dark deuterons have been produced,
larger dark nuclei can be formed by rearrangement collisions,
such as $d_n + d_2 \to d_{n+1} + d$.  The formation of dark deuterons
is a bottleneck that must be overcome by 3-body recombination 
in order to form the larger dark nuclei.
We wish to determine whether this bottleneck can be overcome 
in the early universe when the dark matter is still in thermal equilibrium.
To answer this, we can ignore dark nuclei $d_n$ with $n \ge 3$
and consider only the time evolution for 
dark nucleons and dark deuterons.  The only reactions we need to take into account 
are 3-body recombination and the dark deuteron breakup reaction 
($d_2 +d \leftrightarrow d+d+d $).  We wish to determine whether
a significant population of dark deuterons can be generated in the early universe.

We denote the number densities of the dark nucleon and the dark deuteron 
by $n_1(t)$ and $n_2(t)$. We assume the number densities of 
dark nuclei with larger dark baryon number are negligible, 
so the total dark baryon number density is 
\be
n_\text{dark}(t) =n_1(t) +2 n_2(t).
\label{eq:dn12}
\ee
The time evolution equations for $n_1(t)$ and $n_2(t)$ 
obtained from the Boltzmann equation are
\begin{subequations}
\begin{align}
 \left( \frac{d}{dt} + 3H \right) n_1 &=-2 K_3(T)\, n_1^3 +2 K_2(T)\, n_1 n_2
 -2 K_1(T)\, n_1^2
\,,\label{eq:dn1}\\
 \left( \frac{d}{dt} + 3H \right) n_2 &= K_3(T)\, n_1^3 - K_2(T)\, n_1 n_2 - \Gamma_2\, n_2
\,,\label{eq:dn2}
\end{align}
\label{eq:dn1n2}%
\end{subequations}
where $H$ is the Hubble function, $K_3(T)$, $K_2(T)$,  and $K_1(T)$ are temperature-dependent event rate coefficients,
and $\Gamma_2$ is the dark deuteron decay rate.
The Hubble function $H(t)$ depends on time,
being determined by the scale factor $a(t)$ of the universe: $H=d \ln(a) /dt$.
The rate coefficients are functions of the temperature $T (t)$ of the dark matter, which also depends on time.  

We neglect the  effects of the  annihilation of dark nucleons into ordinary matter.
We therefore set $K_1(T)= 0$ and $\Gamma_2 = 0$ in  the rate equations in Eqs.~\eqref{eq:dn1n2}.
If there were such an annihilation process, it would  decrease $n_1$ through 
annihilation collisions of two dark nucleons
and it would  decrease $n_2$ through the annihilation of the two constituents of the dark deuteron.
The rates for both processes are determined by the same parameter Im[$1/a]$,
which appears as a multiplicative parameter in both $K_1(T)$ and $\Gamma_2$.
When $\Gamma_2$ is much larger than $3 H(t)$,
the number density $n_2$ of dark deuterons decreases exponentially.
Any dimers that have been produced by 3-body recombination would decay quickly on a cosmological time scale.
The net effect is that  $n_2$ would remain essentially 0, 
and the decrease in $n_1$ would be given by the 3-body recombination term in Eq.~\eqref{eq:dn1} only.
Since we ignore the annihilation of dark nucleons, our results for the number density of dark deuterons 
can be interpreted as upper bounds.

If we ignore the annihilation terms in the evolution equations in Eqs.~\eqref{eq:dn1n2}, 
we get a simpler equation for the total dark baryon number density:
\be
 \left( \frac{d}{dt} + 3H(t) \right) n_\text{dark}(t)  = 0 \,.
 \label{eq:dndark}
\ee
Using $dt=H^{-1} d\ln a$, the solution is 
\be
n_\text{dark}(t)  = n_\text{dark}(0) \left(\frac{a(0)}{a(t)}\right)^3\,.
 \label{eq:ndarkt}
\ee
The time evolution of the total dark baryon number density  does not depend on
the dark matter interactions;  it is just diluted by the Hubble expansion. 

It is convenient to use the redshift $z$ as an alternative time variable.
The redshift is related to the scale factor $a$ by $1+z(t)=a(0)/a(t)$.
The solution for the total dark baryon number density
in \eq{ndarkt} can be expressed as
\be
n_\text{dark}(z)  = \frac{\rho_\text{cdm}}{m_\chi} \, (1+z)^3 \,,
 \label{eq:ndarkz}
\ee
where $\rho_\text{cdm} = 2.23\times10^{-30}$~g/cm$^3$ 
is the present average mass density of dark matter in the universe\,\cite{Patrignani:2016xqp}  
and $m_\chi$ is the mass of the dark nucleon.
The Hubble function in terms of redshift is given by
\be
H(z)= H_0 \left[ \Omega_\gamma (1+z)^4+\Omega_m (1+z)^3 +\Omega_\Lambda \right]^{1/2}
\,,
\label{eq:Hz}
\ee
where the Hubble constant is $H_0$ = 67.8 km s$^{-1}$ Mpc$^{-1}$
and the fractions of the critical density of the Universe for CMB photons ($\Omega_\gamma$), matter ($\Omega_m$), and dark energy ($\Omega_\Lambda$) are $5.38\times10^{-5}$, 0.308, and 0.692, respectively\,\cite{Patrignani:2016xqp}.

Since the dark nucleons are nonrelativistic after the decoupling, 
their temperature $T(z)$  is proportional to the square of their
average momentum\,\cite{Armendariz-Picon:2013jej}.
On the other hand, the temperature $T_\gamma(z)$ of the photons is proportional 
to their average momentum.
The Hubble expansion changes the momentum of a particle by a factor of $1+z$.
Thus the two temperatures are different functions of the redshift:
\begin{subequations}
\begin{align}
T(z)  &\approx T(0)\,(1 + z)^2
\,, \label{eq:Tdark}
\\
T_\gamma(z) &\approx T_\mathrm{cmb}\, (1 + z)
\,, \label{eq:Tgam}
\end{align}
\label{eq:Tdarkgam}%
\end{subequations}
where $T(0)$ is the present temperature of dark matter 
that has not been captured by gravitational potential wells 
and $ T_\mathrm{cmb}=2.73$~K is the present temperature of the photons.
At decoupling, the dark matter and ordinary matter are in thermal equilibrium: 
$T(z_\text{dc}) =T_\gamma(z_\text{dc})$, 
where $z_\text{dc}$ is the redshift at decoupling.  We are not displaying the dependence on the Standard Model degrees of freedom in these and the following expressions for the temperature for simplicity.  
The variation due to the three-body parameter is larger than the effects from the decreasing number of relativistic degrees of freedom. 
The dark matter temperature is therefore
\be
T(z)  \approx T_\mathrm{cmb}\,\frac{ (1 + z)^2}{1+z_\text{dc}}\,.
\label{eq:T-z}
\ee
The decoupling redshift can be expressed as $z_\text{dc} \approx T(z_\text{dc})/T_\mathrm{cmb}$.
If the thermal decoupling of dark matter and ordinary matter occurs not long after their chemical decoupling,
the decoupling temperature is given approximately 
by the dark-matter mass multiplied by a constant:  $kT(z_\text{dc}) \approx m_\chi/20$ \cite{Steigman:2012nb}.
The resulting estimate for the decoupling redshift is
\be
1+z_\text{dc}  \approx \frac{m_\chi/20}{kT_\mathrm{cmb}}\,.
\label{eq:zdc}
\ee
The mass fraction of the dark matter in the form of dark deuterons is 
\be
f_2(z)  =2 \, n_2(z)/n_\text{dark}(z) \,. 
 \label{eq:dfraction}
\ee
If we ignore the annihilation terms in the evolution equations in Eqs.~\eqref{eq:dn1n2}, 
the dark deuteron fraction satisfies the differential equation
\be
\frac{d}{dz} f_2 =
\frac{1}{(1+z) H} \Big[- 2\, K_3(T)\, n_\text{dark}^2\, (1-f_2)^3   + K_2(T)\,  n_\text{dark}\, f_2(1-f_2) \Big].
 \label{eq:df2}
\ee
We have used  $dt = -[(1+z)H]^{-1} \,dz$ to replace the time derivative by a redshift derivative.
Given $H(z)$,  $n_\text{dark}(z)$, and  $T(z)$ in \eqss{Hz}{ndarkz}{T-z},
our problem reduces to solving this single differential equation for $f_2(z)$
subject to the initial condition $f_2 (z_\text{dc}) = 0$, where $z_\text{dc}$ is given in \eq{zdc}.

\subsection{Approximation in scaling and threshold regions} 
\label{sec:approx}

The evolution equation for  the dark deuteron fraction  with redshift in Eq.~\eqref{eq:df2}
involves the rate coefficients $K_2(T)$ and $K_3(T)$.
If dark nucleons are identical bosons with a large positive scattering length
and if $kT$ is much smaller than the energy scale $E_0$ set by the range,
the rate coefficients are given in Eqs.~\eqref{eq:K2T} and \eqref{eq:K3T}.
The  rate coefficients have simple behavior in the low-temperature limit $kT \ll E_2$, 
where $E_2= 1/(m_\chi a^2)$ is the dark deuteron binding energy,
and in the scaling region $E_2 \ll kT \ll E_0$, 
where $E_0=1/(m_\chi r_0^2)$ is the energy scale set by the range.
We can use those results to determine the qualitative behavior of the dark deuteron fraction in those  regions.

In the scaling region $E_2 \ll kT \ll E_0$, the rate coefficients $K_2(T)$ and $K_3(T)$ 
have the limiting behaviors given in Eqs.~\eqref{eq:K2scaling} and \eqref{eq:K3scaling}.
They scale as $\lambda_T$ and $\lambda_T^4$, respectively, where $\lambda_T$ is the thermal wavelength,
which is proportional to $(1+z)^{-1}$:
\be
\lambda_T =\left(\frac{2\pi  (1+ z_\text{dc})}{m_\chi \, k T_\text{cmb}} \right)^{1/2} \frac{1}{1+z}
\,.
\label{eq:lamT}
\ee
Since $n_\mathrm{dark}$ is proportional to $(1+z)^3$,
the products $K_2\,  n_\text{dark}$ and $K_3 \, n_\text{dark}^2$ are both proportional to $(1+z)^2$.
Thus there can be an equilibrium value of $f_2$ for which the two terms on the right side of Eq.~\eqref{eq:df2} cancel.
The ratio of $K_3(T)$ and $K_2(T)$ is given in Eq.~\eqref{eq:K3T-K2T}.  The equilibrium fraction satisfies
\be
\frac{f_2}{(1-f_2)^2} =
4 \sqrt{2} \left(\frac{2\pi (1+ z_\text{dc})}{m_\chi \, k T_\text{cmb}} \right)^{3/2} \frac{\rho_\text{cdm}}{m_\chi} \,.
\label{eq:f2equilibrium}
\ee
If  the equilibrium value of $f_2$ is much less than 1, 
it can be approximated by the right side of Eq.~\eqref{eq:f2equilibrium}.
Upon inserting the estimate for the decoupling redshift in Eq.~\eqref{eq:zdc},
the right side reduces to $(\rho_\text{cdm}/m_\chi)/(k T_\text{cmb})^3$ multiplied by a numerical constant.
Since $\rho_\text{cdm}/(k T_\text{cmb})^3  =0.74$~eV, the equilibrium value of $f_2$ is tiny as long as 
$m_\chi$ is orders of magnitude larger than 1~eV.

In the  low-temperature region $kT \ll E_2$, the rate coefficients $K_2(T)$ and $K_3(T)$ 
have the limiting behaviors given in Eqs.~\eqref{eq:K2threshold} and \eqref{eq:K3threshold}.
They are proportional to $\lambda_T^{-6}\, e^{-\lambda_T^2/a^2} $  and $ \lambda_T^{-6} $, respectively.
The dark deuteron breakup is exponentially suppressed by the Boltzmann factor,
so the breakup term  in the rate equation can be dropped. 
If the value of $f_2$ is much less than 1, we need to keep only the leading terms in $f_2$ 
in  the recombination term  in \eq{df2}. 
The rate equation then simplifies to
\be
\frac{d\ }{dz}f_2 = -2K_3(0) \frac{n_\text{dark}^2}{ (1+z)H}.
 \label{eq:df2th}
 \ee
When $z\gg 10^4$, the Hubble function in \eq{Hz} can be approximated as $H(z) \approx H_0 \Omega_\gamma^{1/2} z^2$.
The solution of Eq.~\eqref{eq:df2th}  is then
\be
f_2(z)  = f_2(0)  - C_3(a/a_+)\, \frac{a^4  \rho_\text{cdm}^2}{2H_0 \Omega_\gamma^{1/2} m_\chi^3} z^4\, ,
 \label{eq:f2zero}
\ee
where $f_2(0)$ is the present dark-deuteron fraction for  dark matter 
that has not been captured by gravitational potential wells
and $C_3(a/a_+)$ can be approximated by Eq.~\eqref{eq:C3approx}.
The approach to $f_2(0)$ is predicted to be $z^4$ multiplied by a coefficient whose dependence on $a$
is $C_3(a/a_+)\, a^4$.
The value of $f_2(0)$ should be determined by a boundary condition from the region of larger $z$ 
where $kT(z)$ is comparable to $E_2$. 
We are unable to determine $f_2(0)$ analytically, but it should depend
log-periodically on the three-body parameter $a_+$.
Our numerical results for $f(0)$ are consistent with an expression  linear in $C_3$.

The evolution equation for  the dark deuteron fraction  $f_2(z)$ in Eq.~\eqref{eq:df2}
with the rate coefficients $K_2(T)$ and $K_3(T)$ in Eqs.~\eqref{eq:K2T} and \eqref{eq:K3T}
applies all the way back to the decoupling redshift provided the decoupling temperature
is smaller than the scale $E_0$ set by the range: $kT(z_\text{dc})< E_0$.
If $E_0$ is smaller  than $kT(z_\text{dc})$,
the simple expressions for $K_2(T)$ and $K_3(T)$ in Eqs.~\eqref{eq:K2T} and \eqref{eq:K3T}
are not applicable until $kT$ decreases to below $E_0$.
However once $kT$ enters the scaling region $E_2 \ll kT \ll E_0$,
$f_2$ will be driven quickly to the equilibrium value given by Eq.~\eqref{eq:f2equilibrium}.
Thus  the present value of $f_2$ is completely determined by the scattering length $a$ 
and the 3-body parameter $a_+$ provided only that the decoupling temperature
is much larger than the scale $E_2 = 1/(m_\chi a^2)$.
This condition $kT(z_\text{dc}) \gg E_2$ can be expressed approximately as $m_\chi a \gg \sqrt{20}$.

\subsection{Numerical Results}

Assuming the dark nucleons are identical bosons with a large scattering length,
the few-body parameters are the dark nucleon mass $m_\chi$, the scattering length $a$, 
and the three-body  parameter  $a_+$.
For the mass and the scattering length, we use values that can solve small-scale structure problems of the universe.
The  values that give the best fit to the data points for 
$\langle v\, \sigma_\mathrm{elastic} \rangle$ versus $\langle v  \rangle$ in Figure~\ref{fig:vsigma-v} are
$m_\chi = 19$~GeV and $a=17$~fm.  Since the 3-body parameter $a_+$
 is only defined modulo multiplicative factors of $\lambda_0\approx 22.69$, 
the complete range of possibilities is covered by varying $a_+$ from $a$ to $22.69\,a$.

\begin{figure}[t]
\centering
\includegraphics[width=0.8\linewidth]{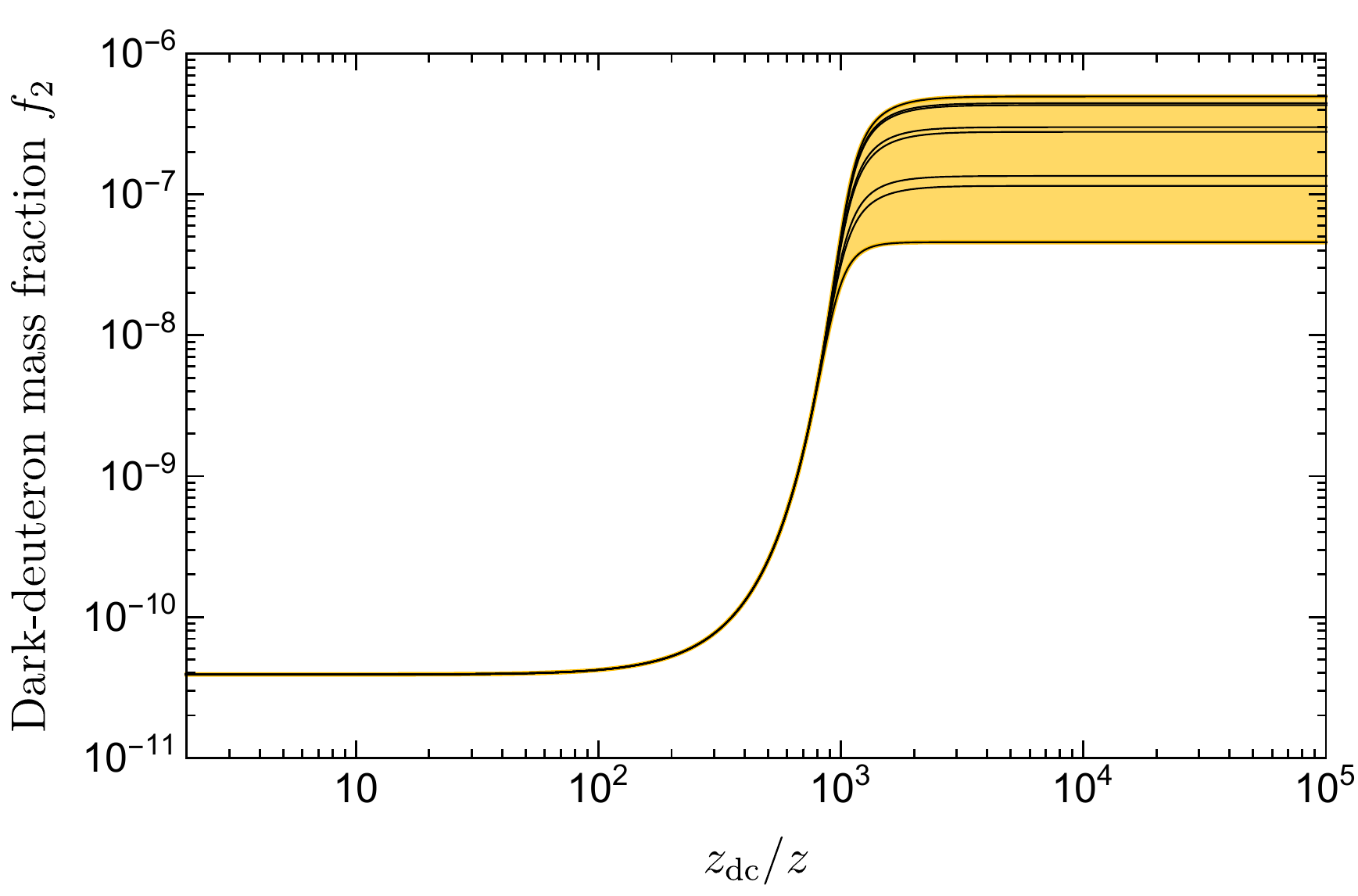}
\caption{Dark-deuteron mass fraction $f_2(z)$ in the early universe as a function of the redshift variable 
$z_\text{dc}/z$ on a log scale
for $m_\chi = 19$~GeV and $a=17$~fm.
The  curves are for 8 values of the 3-body parameter: $a_+/a = \lambda_0^{n/8}$ with $n=0,1,\cdots, 7$.}
\label{fig:fdimer2}
\end{figure}

To determine the dark-deuteron mass fraction $f_2(z)$ as a function of the redshift $z$,
we solve the differential equation in \eq{df2} subject to the initial condition $f_2 (z_\text{dc}) = 0$.
Given the mass $m_\chi = 19$~GeV, the decoupling redshift in Eq.~\eqref{eq:zdc} is $z_\text{dc} \approx 4\times 10^{12}$.
We want to determine whether a significant fraction $f_2$
is ever generated during the subsequent time evolution.

In Figure~\ref{fig:fdimer2}, we show the dark deuteron fraction $f_2$ as a function of 
a red-shift variable $z_\text{dc}/z$ on a log scale.  
This variable increases from 1 at the decoupling time to infinity at the present time.
The band  in Figure~\ref{fig:fdimer2} corresponds to minimizing and maximizing $f_2$ with respect to $a_+$. 
The individual curves are for  eight values of $a_+$
that are equally spaced on a log scale between $a$ and $22.69~a$.
As $z$ decreases from $z_\text{dc}$, the dark deuteron fraction $f_2(z)$ increases 
very  quickly to a plateau of about $4\times 10^{-11}$
from  thermal equilibrium between 3-body recombination and dark deuteron breakup. 
That equilibrium value is consistent with the estimate for the scaling region in \eq{f2equilibrium}. 
We could therefore just as well take the initial value of $f_2$  at the decoupling red shift
to be the equilibrium value given by  Eq.~\eqref{eq:f2equilibrium}.
When the dark matter temperature $T$ decreases 
below the dark deuteron binding energy $1/(m_\chi a^2)=7.1$~keV, 
there is a dramatic increase in $f_2$ by 3 or 4 orders of magnitude.
This feature is expected from the exponential suppression of the breakup process in \eq{K2threshold}
and from the $z^4$ dependence at late times that is predicted by \eq{f2zero}. 
The dark-deuteron fraction plateaus at a value $f_2(0)$ that depends log-periodically on the 3-body parameter $a_+$
and can be approximated by
\be
f_2(0)  = (4.6\times 10^{-8})+ (6.7\times 10^{-9}) \, C_3(a/a_+).
 \label{eq:f2zero-C3}
\ee
The maximum value of $f_2(0)$ from varying $a_+$ is larger than the minimum value by a factor of  11.
The fraction $f_2(0)$ has its minimum when the value of $a_+/a$ is just a few percent lower than 1 
(or equivalently $\lambda_0=22.69$),
which is the value for which there is total destructive interference in the 3-body recombination rate at  zero temperature.
It has its maximum when the value of $a_+/a$ is just a few percent lower than $\lambda_0^{1/2}=4.76$.

\begin{figure}[t]
\centering
\includegraphics[width=0.8\linewidth]{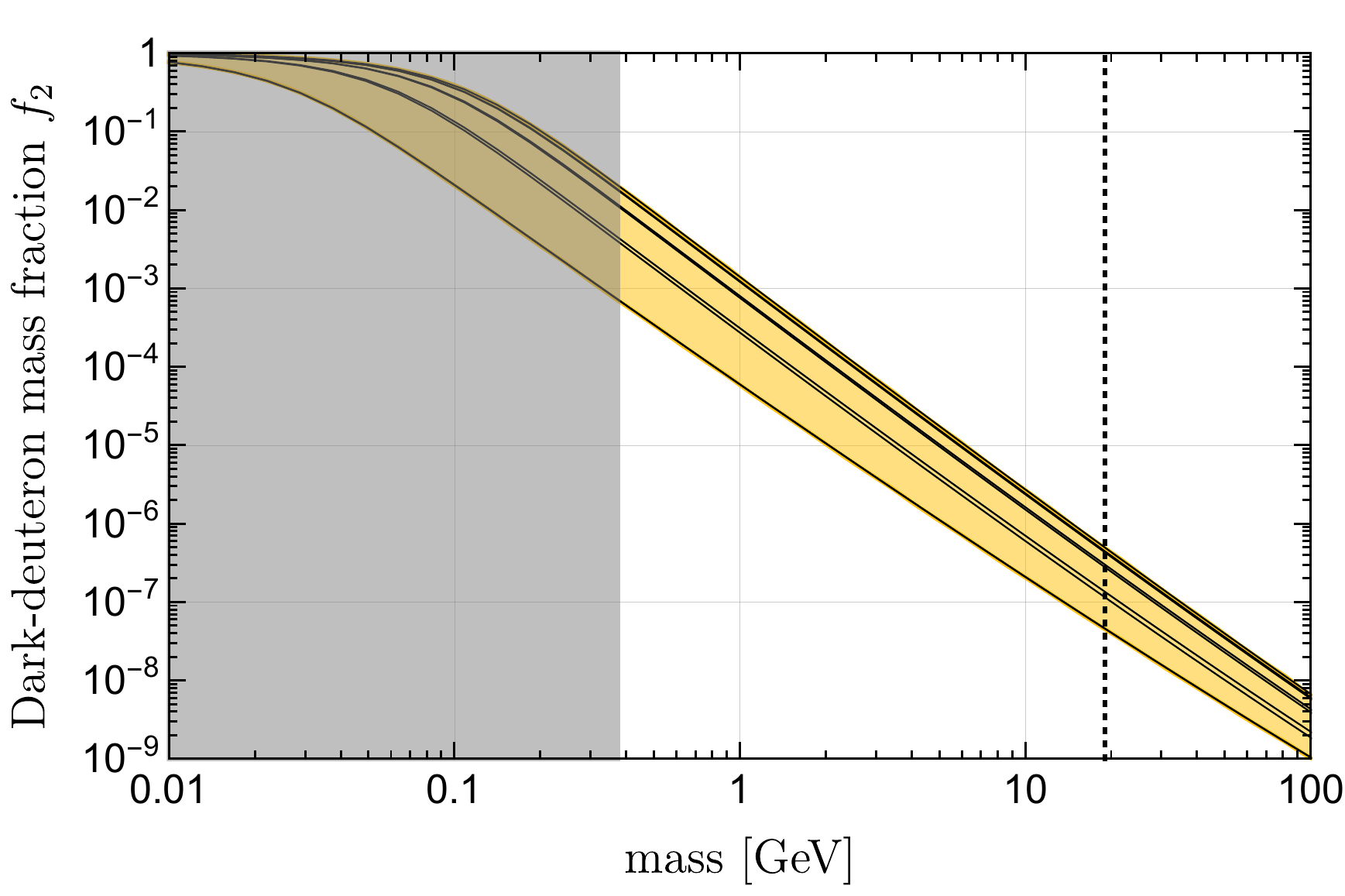}
\caption{
Dark-deuteron mass fraction $f_2(z=0)$ at late times as a function of the dark matter mass $m_\chi$
with the scattering length $a$ determined by $\sigma_\text{elastic}/ m_\chi = 2~\cm^2/\gram$.
The  curves are for 8 values of the 3-body parameter: $a_+/a = \lambda_0^{n/8}$ with $n=0,1,\cdots, 7$.
The vertical dotted line marks the mass  $m_\chi = 19$~GeV used in Figure~\ref{fig:fdimer2}.
The results are reliable only well outside the shaded region where $m_\chi a \gg\sqrt{20}$.
}
\label{fig:f2_mass}
\end{figure}

The results for $f_2(z)$ shown  in Figure~\ref{fig:fdimer2} are for parameters $m_\chi = 19$~GeV and $a=17$~fm
that solve small-scale structure problems of the universe, as illustrated in Figure~\ref{fig:vsigma-v}.
However there are also mechanisms involving baryonic physics that can solve or at least ameliorate
the small-scale structure problems.
We first relax the constraint on $m_\chi$ and $a$ by ignoring the results for $\langle v\, \sigma_\mathrm{elastic}\rangle$ 
versus  $\langle v\rangle$ from clusters of galaxies. 
The results in Figure~\ref{fig:vsigma-v} from galaxies only are roughly compatible 
with a  cross section that at low velocities approaches $\sigma_\mathrm{elastic}/m =2~\mathrm{cm}^2$/g.
This requires the constraint  $8 \pi a^2/m_\chi = 2~ \mathrm{cm}^2$/g.
Given $m_\chi$, the scattering length $a$ is determined.
The results for the dark deuteron fraction $f_2(0)$ at late times as a function of $m_\chi$ are shown in Figure~\ref{fig:f2_mass}.
If $m_\chi$ is too small, the decoupling temperature $kT(z_\text{dc}) \approx m_\chi /20$
cannot be much larger than the binding energy $E_2= 1/(m_\chi a^2)$.
In this case, the temperatures after decoupling do not include  a scaling region in which $T \gg E_2$,
so $f_2$ is not determined by $a$ and $a_+$ only, but has additional sensitivity to the range $r_0$.
This additional sensitivity to $r_0$  is avoided if $m_\chi a \gg \sqrt{20}$, which implies $m_\chi \gg 0.4$~GeV.  
This requires $m_\chi$ to be well above the shaded region in Figure~\ref{fig:f2_mass}. 
In the unshaded region,
the fraction $f_2(0)$ scales  roughly as $m_\chi^{-2.5}$. 
At $m_\chi = 1$~GeV, the range of $f_2(0)$ from varying $a_+$ is from $6\times10^{-5}$ to $2\times 10^{-3}$.
Thus a dark deuteron fraction as large as $10^{-3}$ is possible if the dark-matter elastic cross section 
at low velocities is $2~\mathrm{cm}^2$/g.

If the small-scale structure problems of galaxies are ameliorated by mechanisms involving baryonic physics,
the constraint on $m_\chi$ and $a$ becomes
the inequality $8 \pi a^2/m_\chi < 2~ \mathrm{cm}^2$/g.
At a given value of $m_\chi$, this allows the scattering length to  be decreased.
This can only decrease the dark deuteron fraction $f_2(0)$ at late times.

\begin{figure}[t]
\centering
\includegraphics[width=0.8\linewidth]{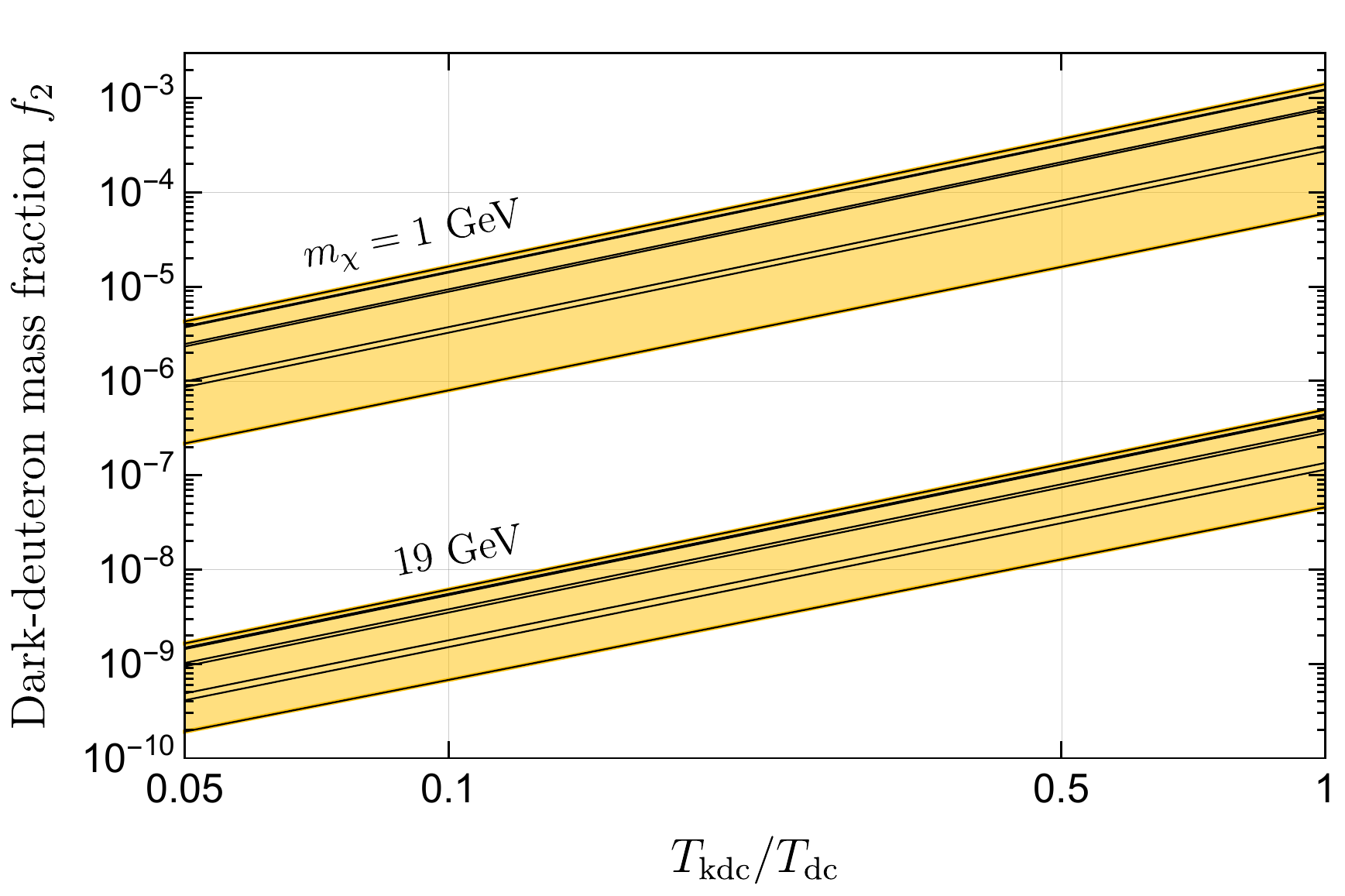}
\caption{
Dark-deuteron mass fraction $f_2(z=0)$ at late times as a function of 
ratio  $T_\text{kdc}/T_\text{dc}$ of the temperatures for kinetic decoupling
and chemical decoupling for the masses 1~GeV (upper band) and 19~GeV (lower band). 
The  curves are for 8 values of the 3-body parameter: $a_+/a = \lambda_0^{n/8}$ with $n=0,1,\cdots, 7$.
 }
\label{fig:f2_T}
\end{figure}

The results presented above assume  the kinetic decoupling temperature $T_ \text{kdc}$ 
is very close to the chemical decoupling temperature $T_ \text{dc}$ \cite{Binder:2017rgn}.   
We now consider the case when $T_ \text{kdc}$ is significantly smaller than $T_ \text{dc}$.  
The value of $T_\text{kdc}$  depends on
the interactions between dark matter and ordinary matter.
We treat it as an unknown parameter
and  simply describe how $f_2(0)$ scales with $T_\text{kdc}/T_\text{dc}$.
During thermal equilibrium, the dark matter temperature is the same as the photon temperature  in \eq{Tgam}, 
while after kinetic decoupling the temperature is quadratic in $z$ as in \eq{Tdark}.
The dark matter temperature can be written as
\bse
\bqa
T(z) &\approx& (1+z)T_\text{cmb}~~~~~~~~~~~z>z_\text{kdc} ,
\\
&\approx& 
\left(\frac{1+z}{1+z_\text{kdc}}\right)^2 \, T_\text{kdc}~~~z<z_\text{kdc},
\eqa
\label{eq:T-z-kin}%
\ese
where $z_\text{kdc} \approx T_\text{kdc}/T_\text{cmb}-1$ is the redshift at the kinetic decoupling temperature. 
By inserting \eq{T-z-kin} into \eq{df2} instead of \eq{T-z}, we obtain the fraction with kinetic decoupling taken 
into account. 
Figure~\ref{fig:f2_T} shows the fraction $f_2(z=0)$ at late times as a function of $T_\text{kdc}/T_\text{dc}$ 
for the masses 1~GeV (upper band) and 19~GeV (lower band). 
As shown in Figure~\ref{fig:f2_T}, as $T_\text{kdc}$ decreases, the fraction decreases,
scaling roughly as $(T_\text{kdc}/T_\text{dc})^{1.9}$.

\section{Discussion and conclusion}
\label{sec:discussion}

 The predictions of $\Lambda$CDM cosmology face a number of challenges at small scales.  
 These small-scale structure problems can be resolved either by effects of baryons on structure formation 
 or by novel dark matter dynamics.  Self-interacting dark matter is a paradigm that can solve the small-scale 
 structure problems in $\Lambda$CDM cosmology while remaining consistent with other cosmological data.  
Perhaps the most predictive model of self-interacting dark matter involves a near-threshold S-wave resonance that produces a
large  scattering length\,\cite{Braaten:2013tza}.  This near-threshold resonance is a bound state if the scattering length is positive.  The signatures for this two-body bound state (darkonium or dark deuteron)  in direct detection and directional detection experiments have been studied \cite{Laha:2013gva, Laha:2015yoa}.   
In this paper, we have studied the production rate of the dark deuteron in the early universe.

We first compared the predictions of the near-threshold S-wave resonance model 
with cross sections for self-interacting dark matter in different astrophysical objects
determined by Kaplinghat, Tulin and Yu \cite{Kaplinghat:2015aga}.
They showed that a dark-photon model with three adjustable parameters 
can reproduce the velocity dependence of the self-interaction cross section, which spans two orders of magnitude in velocity \cite{Kaplinghat:2015aga}.  We find that the near-threshold S-wave resonance model provides an equally good fit to these astrophysical data (see Fig.\,\ref{fig:vsigma-v}) with only two adjustable parameters:
the mass $m_\chi$ of the dark-matter particle and the scattering length $a$.
The best-fit values are $m_\chi = 19$~GeV and $a=\pm 17$~fm.

We have assumed the dark nucleons are identical bosons with a large positive scattering length.  
The smallest universal bound cluster is the dark deuteron $d_2$.
The simplest reaction that can form this bound cluster is 
3-body recombination into  the dark deuteron: $d + d + d   \to d_2 + d$.
The three-body recombination rate is a function of the mass $m_\chi$, the scattering length $a$, 
and a three-body parameter $a_+$, with the dependence on $a_+$ being log-periodic with discrete scaling factor 22.7.
If the temperature at decoupling is much larger than the binding energy of the dark deuteron, the present fraction $f_2(0)$
of dark matter in the form of dark deuterons is completely determined by these three parameters.
For $m_\chi = 19$~GeV and $a=17$~fm, the fraction $f_2(z)$ at early red shifts
has an equilibrium value of about $4 \times 10^{-11}$. When the dark-matter temperature decreases
to below the binding energy of the dark deuteron, which occurs at a red shift  $z \approx 10^{10}$, $f_2(z)$
increases by orders of magnitude to a value between $4 \times 10^{-8}$ 
and $5 \times 10^{-7}$ that depends on $a_+$.   

The present fraction $f_2(0)$ of dark matter in the form of dark deuterons can be increased
 by relaxing the constraint on $m_\chi$ and $a$ from solving small-scale structure problems 
and decreasing $m_\chi$. 
However  the decoupling temperature must be much larger than the binding energy of the dark deuteron
for $f_2(0)$ to be insensitive to the range of the interactions. Given this constraint,
$f_2(0)$ cannot be larger than about $10^{-3}$.
If the system remains in thermal equilibrium longer after chemical decoupling, the fraction $f_2(0)$ decreases, 
scaling approximately as the 2nd power of the ratio of the temperatures for  kinetic and chemical decoupling.
We conclude that
a significant population of dark deuterons cannot be produced in the early universe by  3-body recombination
of dark matter particles with a large scattering length.
Since the production of dark deuterons is a bottleneck for the formation of larger bound clusters,
we conclude that the formation of bound clusters in the early universe would require additional microphysics.
An example is a light mediator that allows radiative fusion reactions.

If the large scattering length $a$ is negative,
the smallest universal bound clusters are Efimov clusters $d_3$ (``dark tritons'').
The simplest reaction that can form bound clusters is 
4-body recombination into a dark triton: $d + d + d + d  \to d_3 + d$.
The rate for 4-body recombination is suppressed compared to the rate for 3-body recombination 
by an additional factor of the number density of dark matter particles.
Since a significant population of dark deuterons cannot be produced in the early universe
by  3-body recombination,  a significant population of dark tritons cannot be produced
by  4-body recombination either.
Since the production of dark tritons is a bottleneck for the production of larger dark nuclei,
a significant number of dark nuclei will not be formed in the early universe 
if the dark nucleons are identical bosons with a large negative scattering length.

Identical bosons are not the only types of particles for which there is
dramatic enhancement of the 3-body recombination rate at low temperature
when the scattering length is large.
The degree to which 3-body recombination is enhanced
depends on the symmetries and mass ratios of the particles with large scattering lengths.
Three-body recombination requires the three particles to come within a distance of order 
the de Broglie wavelength of the final-state particles, which is of order $1/a$ if the collision energy is small.
For identical bosons,  the 3-body recombination rate $K_3(T)$ in the low-temperature
limit is proportional to $a^4$.  If the dark matter consists of the two spin states of a spin-$\frac12$ fermion,
$K_3(T\to 0)$ is suppressed by $(r_0/a)^2$, where $r_0$ is the range,
because the Pauli exclusion principle suppresses the contribution from the region where the separations of the  three fermions
are all of order $a$. If the dark matter consists of the four spin states of two spin-$\frac12$ fermions,
there is no such suppression and $K_3(T\to 0)$ is proportional to $a^4$.

We can also show that a significant fraction of  dark deuterons 
cannot form once the dark matter particles fall inside the gravitational potential well of a galaxy.  
It is easy to put an upper bound on the rate of increase in the dark deuteron fraction in the Milky Way 
from 3-body recombination.  The maximum possible rate of increase in $n_2$ is $(67.1\, a^4/m_\chi)n_1^3$.
The dark matter mass density in the solar system, which is about 8 kpc from
the center of the Milky Way, is $m_\chi n_1 = 0.3$~GeV/cm$^3$.
If feedback between strongly interacting dark matter and baryons is taken into account,
the radius of the dark matter core of the Milky Way may be about 0.3 kpc \cite{Kaplinghat:2013xca}.
The dark matter mass density in the core of the Milky Way may be  about $m_\chi n_1 = 8$~GeV/cm$^3$.
For $m_\chi = 19$~GeV and $a=17$~fm, the maximum rate of increase of $f_2$ is about $10^{-51}$/s.
The age of the Milky Way is about $10~\text{Gyr}\approx 3\times10^{17}~\text{s}$, 
so we see that the dark deuteron fraction remains negligible.
If we relax the constraints on $m_\chi$ and $a$ from solving small-scale structure problems
but keep the binding energy of the dark deuteron small compared to the decoupling temperature,
the rate of increase of $f_2$ can be made larger at most by about an order of magnitude.
Dwarf galaxies can have higher dark matter densities than the Milky Way, 
but the rate  of increase in the dark deuteron fraction is small in those systems too.

Although 3-body recombination of dark matter particles  is unable to build up 
a large fraction of dark deuterons in the early universe, it may still have a significant effect on dark matter annihilation.
If a pair of dark matter particles has an annihilation scattering channel, the constituents of a dark deuteron will 
eventually annihilate once the dark deuteron is formed.
Three-body recombination therefore provides an additional  annihilation channel.
If the dark matter particles have a large scattering length $a$, the annihilation scattering cross section
and the dark deuteron decay rate are both determined by $m_\chi$ and $a$ up to a multiplicative constant
that cancels in their ratio.
The resonant enhancement of annihilation scattering  can induce a second period 
of dark matter annihilation  after the thermal freezeout \cite{Binder:2017lkj}.
The effects of reannihilation have been explored in a dark photon model  \cite{Binder:2017lkj}.
A near-threshold S-wave resonance model provides a more predictive framework 
in which the effects of reannihilation through 3-body recombination can also be easily explored.

\begin{acknowledgments}
We thank M.~Kaplinghat, S.~Tulin, and H.-B.~Yu for providing their data in Figure~\ref{fig:vsigma-v}
and M.~Kaplinghat for valuable comments.
This research of E.B.\ was supported in part by the
Department of Energy under the grant DE-SC0011726
and by the National Science Foundation under grant PHY-1607190.
D.K. is supported by NSFC through Grant No. 11875112.
R.L. is supported by German Research  Foundation  (DFG)  under  Grant  Nos.  EXC-1098, KO 4820/1-1, FOR 2239, 
and from the European Research Council (ERC) under the European Union's Horizon 2020 research and innovation 
programme  (grant  agreement  No.  637506,  ``$\nu$Directions") awarded to Joachim Koop.
We acknowledge the INT program ``Multi-Scale Problems Using Effective Field Theories (INT-18-1b)", 
during which this work was finalized.
D.K.\ would like to thank the hospitality of High Energy Theory Group, Ohio State University, 
where an important part of this work was carried out.
\end{acknowledgments}

\bibliographystyle{JHEP}
\bibliography{references}	

\providecommand{\href}[2]{#2}\begingroup\raggedright\begin{thebibliography}{10}

\bibitem{Braaten:2013tza}
E.~Braaten and H.~W. Hammer, \emph{{Universal Two-body Physics in Dark Matter
  near an S-wave Resonance}},
  \href{https://doi.org/10.1103/PhysRevD.88.063511}{\emph{Phys. Rev.}
  {\bfseries D88} (2013) 063511},
  [\href{https://arxiv.org/abs/1303.4682}{{\ttfamily 1303.4682}}].

\bibitem{Laha:2013gva}
R.~Laha and E.~Braaten, \emph{{Direct detection of dark matter in universal
  bound states}}, \href{https://doi.org/10.1103/PhysRevD.89.103510}{\emph{Phys.
  Rev.} {\bfseries D89} (2014) 103510},
  [\href{https://arxiv.org/abs/1311.6386}{{\ttfamily 1311.6386}}].

\bibitem{Laha:2015yoa}
R.~Laha, \emph{{Directional detection of dark matter in universal bound
  states}}, \href{https://doi.org/10.1103/PhysRevD.92.083509}{\emph{Phys. Rev.}
  {\bfseries D92} (2015) 083509},
  [\href{https://arxiv.org/abs/1505.02772}{{\ttfamily 1505.02772}}].

\bibitem{Shepherd:2009sa}
W.~Shepherd, T.~M.~P. Tait and G.~Zaharijas, \emph{{Bound states of weakly
  interacting dark matter}},
  \href{https://doi.org/10.1103/PhysRevD.79.055022}{\emph{Phys. Rev.}
  {\bfseries D79} (2009) 055022},
  [\href{https://arxiv.org/abs/0901.2125}{{\ttfamily 0901.2125}}].

\bibitem{Khlopov:2010pq}
M.~{\relax Yu}. Khlopov, A.~G. Mayorov and E.~{\relax Yu}. Soldatov,
  \emph{{Composite Dark Matter and Puzzles of Dark Matter Searches}},
  \href{https://doi.org/10.1142/S0218271810017962}{\emph{Int. J. Mod. Phys.}
  {\bfseries D19} (2010) 1385--1395},
  [\href{https://arxiv.org/abs/1003.1144}{{\ttfamily 1003.1144}}].

\bibitem{Cline:2013zca}
J.~M. Cline, Z.~Liu, G.~Moore and W.~Xue, \emph{{Composite strongly interacting
  dark matter}}, \href{https://doi.org/10.1103/PhysRevD.90.015023}{\emph{Phys.
  Rev.} {\bfseries D90} (2014) 015023},
  [\href{https://arxiv.org/abs/1312.3325}{{\ttfamily 1312.3325}}].

\bibitem{Foot:2014mia}
R.~Foot, \emph{{Mirror dark matter: Cosmology, galaxy structure and direct
  detection}}, \href{https://doi.org/10.1142/S0217751X14300130}{\emph{Int. J.
  Mod. Phys.} {\bfseries A29} (2014) 1430013},
  [\href{https://arxiv.org/abs/1401.3965}{{\ttfamily 1401.3965}}].

\bibitem{Petraki:2014uza}
K.~Petraki, L.~Pearce and A.~Kusenko, \emph{{Self-interacting asymmetric dark
  matter coupled to a light massive dark photon}},
  \href{https://doi.org/10.1088/1475-7516/2014/07/039}{\emph{JCAP} {\bfseries
  1407} (2014) 039}, [\href{https://arxiv.org/abs/1403.1077}{{\ttfamily
  1403.1077}}].

\bibitem{Detmold:2014qqa}
W.~Detmold, M.~McCullough and A.~Pochinsky, \emph{{Dark Nuclei I: Cosmology and
  Indirect Detection}},
  \href{https://doi.org/10.1103/PhysRevD.90.115013}{\emph{Phys. Rev.}
  {\bfseries D90} (2014) 115013},
  [\href{https://arxiv.org/abs/1406.2276}{{\ttfamily 1406.2276}}].

\bibitem{Detmold:2014kba}
W.~Detmold, M.~McCullough and A.~Pochinsky, \emph{{Dark nuclei. II. Nuclear
  spectroscopy in two-color QCD}},
  \href{https://doi.org/10.1103/PhysRevD.90.114506}{\emph{Phys. Rev.}
  {\bfseries D90} (2014) 114506},
  [\href{https://arxiv.org/abs/1406.4116}{{\ttfamily 1406.4116}}].

\bibitem{Wise:2014jva}
M.~B. Wise and Y.~Zhang, \emph{{Stable Bound States of Asymmetric Dark
  Matter}}, \href{https://doi.org/10.1103/PhysRevD.90.055030,
  10.1103/PhysRevD.91.039907}{\emph{Phys. Rev.} {\bfseries D90} (2014) 055030},
  [\href{https://arxiv.org/abs/1407.4121}{{\ttfamily 1407.4121}}].

\bibitem{vonHarling:2014kha}
B.~von Harling and K.~Petraki, \emph{{Bound-state formation for thermal relic
  dark matter and unitarity}},
  \href{https://doi.org/10.1088/1475-7516/2014/12/033}{\emph{JCAP} {\bfseries
  1412} (2014) 033}, [\href{https://arxiv.org/abs/1407.7874}{{\ttfamily
  1407.7874}}].

\bibitem{Wise:2014ola}
M.~B. Wise and Y.~Zhang, \emph{{Yukawa Bound States of a Large Number of
  Fermions}}, \href{https://doi.org/10.1007/JHEP10(2015)165,
  10.1007/JHEP02(2015)023}{\emph{JHEP} {\bfseries 02} (2015) 023},
  [\href{https://arxiv.org/abs/1411.1772}{{\ttfamily 1411.1772}}].

\bibitem{Hardy:2014mqa}
E.~Hardy, R.~Lasenby, J.~March-Russell and S.~M. West, \emph{{Big Bang
  Synthesis of Nuclear Dark Matter}},
  \href{https://doi.org/10.1007/JHEP06(2015)011}{\emph{JHEP} {\bfseries 06}
  (2015) 011}, [\href{https://arxiv.org/abs/1411.3739}{{\ttfamily 1411.3739}}].

\bibitem{Appelquist:2015yfa}
T.~Appelquist et~al., \emph{{Stealth Dark Matter: Dark scalar baryons through
  the Higgs portal}},
  \href{https://doi.org/10.1103/PhysRevD.92.075030}{\emph{Phys. Rev.}
  {\bfseries D92} (2015) 075030},
  [\href{https://arxiv.org/abs/1503.04203}{{\ttfamily 1503.04203}}].

\bibitem{Hardy:2015boa}
E.~Hardy, R.~Lasenby, J.~March-Russell and S.~M. West, \emph{{Signatures of
  Large Composite Dark Matter States}},
  \href{https://doi.org/10.1007/JHEP07(2015)133}{\emph{JHEP} {\bfseries 07}
  (2015) 133}, [\href{https://arxiv.org/abs/1504.05419}{{\ttfamily
  1504.05419}}].

\bibitem{Petraki:2015hla}
K.~Petraki, M.~Postma and M.~Wiechers, \emph{{Dark-matter bound states from
  Feynman diagrams}},
  \href{https://doi.org/10.1007/JHEP06(2015)128}{\emph{JHEP} {\bfseries 06}
  (2015) 128}, [\href{https://arxiv.org/abs/1505.00109}{{\ttfamily
  1505.00109}}].

\bibitem{An:2015pva}
H.~An, B.~Echenard, M.~Pospelov and Y.~Zhang, \emph{{Probing the Dark Sector
  with Dark Matter Bound States}},
  \href{https://doi.org/10.1103/PhysRevLett.116.151801}{\emph{Phys. Rev. Lett.}
  {\bfseries 116} (2016) 151801},
  [\href{https://arxiv.org/abs/1510.05020}{{\ttfamily 1510.05020}}].

\bibitem{Tsai:2015ugz}
Y.~Tsai, L.-T. Wang and Y.~Zhao, \emph{{Dark Matter Annihilation Decay at The
  LHC}}, \href{https://doi.org/10.1103/PhysRevD.93.035024}{\emph{Phys. Rev.}
  {\bfseries D93} (2016) 035024},
  [\href{https://arxiv.org/abs/1511.07433}{{\ttfamily 1511.07433}}].

\bibitem{Bi:2016gca}
X.-J. Bi, Z.~Kang, P.~Ko, J.~Li and T.~Li, \emph{{Asymmetric Dark Matter Bound
  State}}, \href{https://doi.org/10.1103/PhysRevD.95.043540}{\emph{Phys. Rev.}
  {\bfseries D95} (2017) 043540},
  [\href{https://arxiv.org/abs/1602.08816}{{\ttfamily 1602.08816}}].

\bibitem{Kribs:2016cew}
G.~D. Kribs and E.~T. Neil, \emph{{Review of strongly-coupled composite dark
  matter models and lattice simulations}},
  \href{https://doi.org/10.1142/S0217751X16430041}{\emph{Int. J. Mod. Phys.}
  {\bfseries A31} (2016) 1643004},
  [\href{https://arxiv.org/abs/1604.04627}{{\ttfamily 1604.04627}}].

\bibitem{Kouvaris:2016ltf}
C.~Kouvaris, K.~Lang{\ae}ble and N.~G. Nielsen, \emph{{The Spectrum of
  Darkonium in the Sun}},
  \href{https://doi.org/10.1088/1475-7516/2016/10/012}{\emph{JCAP} {\bfseries
  1610} (2016) 012}, [\href{https://arxiv.org/abs/1607.00374}{{\ttfamily
  1607.00374}}].

\bibitem{Nozzoli:2016coi}
F.~Nozzoli, \emph{{A balance for Dark Matter bound states}},
  \href{https://doi.org/10.1016/j.astropartphys.2017.03.005}{\emph{Astropart.
  Phys.} {\bfseries 91} (2017) 22--33},
  [\href{https://arxiv.org/abs/1608.00405}{{\ttfamily 1608.00405}}].

\bibitem{Butcher:2016hic}
A.~Butcher, R.~Kirk, J.~Monroe and S.~M. West, \emph{{Can Tonne-Scale Direct
  Detection Experiments Discover Nuclear Dark Matter?}},
  \href{https://doi.org/10.1088/1475-7516/2017/10/035}{\emph{JCAP} {\bfseries
  1710} (2017) 035}, [\href{https://arxiv.org/abs/1610.01840}{{\ttfamily
  1610.01840}}].

\bibitem{Asadi:2016ybp}
P.~Asadi, M.~Baumgart, P.~J. Fitzpatrick, E.~Krupczak and T.~R. Slatyer,
  \emph{{Capture and Decay of Electroweak WIMPonium}},
  \href{https://doi.org/10.1088/1475-7516/2017/02/005}{\emph{JCAP} {\bfseries
  1702} (2017) 005}, [\href{https://arxiv.org/abs/1610.07617}{{\ttfamily
  1610.07617}}].

\bibitem{Petraki:2016cnz}
K.~Petraki, M.~Postma and J.~de~Vries, \emph{{Radiative bound-state-formation
  cross-sections for dark matter interacting via a Yukawa potential}},
  \href{https://doi.org/10.1007/JHEP04(2017)077}{\emph{JHEP} {\bfseries 04}
  (2017) 077}, [\href{https://arxiv.org/abs/1611.01394}{{\ttfamily
  1611.01394}}].

\bibitem{Cirelli:2016rnw}
M.~Cirelli, P.~Panci, K.~Petraki, F.~Sala and M.~Taoso, \emph{{Dark Matter's
  secret liaisons: phenomenology of a dark U(1) sector with bound states}},
  \href{https://doi.org/10.1088/1475-7516/2017/05/036}{\emph{JCAP} {\bfseries
  1705} (2017) 036}, [\href{https://arxiv.org/abs/1612.07295}{{\ttfamily
  1612.07295}}].

\bibitem{Lonsdale:2017mzg}
S.~J. Lonsdale, M.~Schroor and R.~R. Volkas, \emph{{Asymmetric Dark Matter and
  the hadronic spectra of hidden QCD}},
  \href{https://doi.org/10.1103/PhysRevD.96.055027}{\emph{Phys. Rev.}
  {\bfseries D96} (2017) 055027},
  [\href{https://arxiv.org/abs/1704.05213}{{\ttfamily 1704.05213}}].

\bibitem{Gresham:2017zqi}
M.~I. Gresham, H.~K. Lou and K.~M. Zurek, \emph{{Nuclear Structure of Bound
  States of Asymmetric Dark Matter}},
  \href{https://doi.org/10.1103/PhysRevD.96.096012}{\emph{Phys. Rev.}
  {\bfseries D96} (2017) 096012},
  [\href{https://arxiv.org/abs/1707.02313}{{\ttfamily 1707.02313}}].

\bibitem{Mitridate:2017oky}
A.~Mitridate, M.~Redi, J.~Smirnov and A.~Strumia, \emph{{Dark Matter as a
  weakly coupled Dark Baryon}},
  \href{https://doi.org/10.1007/JHEP10(2017)210}{\emph{JHEP} {\bfseries 10}
  (2017) 210}, [\href{https://arxiv.org/abs/1707.05380}{{\ttfamily
  1707.05380}}].

\bibitem{Elor:2018xku}
G.~Elor, H.~Liu, T.~R. Slatyer and Y.~Soreq, \emph{{Complementarity for Dark
  Sector Bound States}},  \href{https://arxiv.org/abs/1801.07723}{{\ttfamily
  1801.07723}}.

\bibitem{Harz:2018csl}
J.~Harz and K.~Petraki, \emph{{Radiative bound-state formation in unbroken
  perturbative non-Abelian theories and implications for dark matter}},
  \href{https://arxiv.org/abs/1805.01200}{{\ttfamily 1805.01200}}.

\bibitem{Biondini:2018pwp}
S.~Biondini and M.~Laine, \emph{{Thermal dark matter co-annihilating with a
  strongly interacting scalar}},
  \href{https://doi.org/10.1007/JHEP04(2018)072}{\emph{JHEP} {\bfseries 04}
  (2018) 072}, [\href{https://arxiv.org/abs/1801.05821}{{\ttfamily
  1801.05821}}].

\bibitem{Biondini:2018xor}
S.~Biondini, \emph{{Bound-state effects for dark matter with Higgs-like
  mediators}}, \href{https://doi.org/10.1007/JHEP06(2018)104}{\emph{JHEP}
  {\bfseries 06} (2018) 104},
  [\href{https://arxiv.org/abs/1805.00353}{{\ttfamily 1805.00353}}].

\bibitem{Cyburt:2015mya}
R.~H. Cyburt, B.~D. Fields, K.~A. Olive and T.-H. Yeh, \emph{{Big Bang
  Nucleosynthesis: 2015}},
  \href{https://doi.org/10.1103/RevModPhys.88.015004}{\emph{Rev. Mod. Phys.}
  {\bfseries 88} (2016) 015004},
  [\href{https://arxiv.org/abs/1505.01076}{{\ttfamily 1505.01076}}].

\bibitem{Consiglio:2017pot}
R.~Consiglio, P.~F. de~Salas, G.~Mangano, G.~Miele, S.~Pastor and O.~Pisanti,
  \emph{{PArthENoPE reloaded}},
  \href{https://arxiv.org/abs/1712.04378}{{\ttfamily 1712.04378}}.

\bibitem{Krnjaic:2014xza}
G.~Krnjaic and K.~Sigurdson, \emph{{Big Bang Darkleosynthesis}},
  \href{https://doi.org/10.1016/j.physletb.2015.11.001}{\emph{Phys. Lett.}
  {\bfseries B751} (2015) 464--468},
  [\href{https://arxiv.org/abs/1406.1171}{{\ttfamily 1406.1171}}].

\bibitem{Braaten:2004rn}
E.~Braaten and H.~W. Hammer, \emph{{Universality in few-body systems with large
  scattering length}},
  \href{https://doi.org/10.1016/j.physrep.2006.03.001}{\emph{Phys. Rept.}
  {\bfseries 428} (2006) 259--390},
  [\href{https://arxiv.org/abs/cond-mat/0410417}{{\ttfamily
  cond-mat/0410417}}].

\bibitem{Weinberg:2013aya}
D.~H. Weinberg, J.~S. Bullock, F.~Governato, R.~Kuzio~de Naray and A.~H.~G.
  Peter, \emph{{Cold dark matter: controversies on small scales}},
  \href{https://doi.org/10.1073/pnas.1308716112}{\emph{Proc. Nat. Acad. Sci.}
  {\bfseries 112} (2014) 12249--12255},
  [\href{https://arxiv.org/abs/1306.0913}{{\ttfamily 1306.0913}}].

\bibitem{Brooks:2014qya}
A.~Brooks, \emph{{Re-Examining Astrophysical Constraints on the Dark Matter
  Model}}, \href{https://doi.org/10.1002/andp.201400068}{\emph{Annalen Phys.}
  {\bfseries 526} (2014) 294--308},
  [\href{https://arxiv.org/abs/1407.7544}{{\ttfamily 1407.7544}}].

\bibitem{Pontzen:2014lma}
A.~Pontzen and F.~Governato, \emph{{Cold dark matter heats up}},
  \href{https://doi.org/10.1038/nature12953}{\emph{Nature} {\bfseries 506}
  (2014) 171--178}, [\href{https://arxiv.org/abs/1402.1764}{{\ttfamily
  1402.1764}}].

\bibitem{DelPopolo:2016emo}
A.~Del~Popolo and M.~Le~Delliou, \emph{{Small scale problems of the
  $\Lambda$CDM model: a short review}},
  \href{https://doi.org/10.3390/galaxies5010017}{\emph{Galaxies} {\bfseries 5}
  (2017) 17}, [\href{https://arxiv.org/abs/1606.07790}{{\ttfamily
  1606.07790}}].

\bibitem{Bullock:2017xww}
J.~S. Bullock and M.~Boylan-Kolchin, \emph{{Small-Scale Challenges to the
  $\Lambda$CDM Paradigm}},
  \href{https://doi.org/10.1146/annurev-astro-091916-055313}{\emph{Ann. Rev.
  Astron. Astrophys.} {\bfseries 55} (2017) 343--387},
  [\href{https://arxiv.org/abs/1707.04256}{{\ttfamily 1707.04256}}].

\bibitem{Buckley:2017ijx}
M.~R. Buckley and A.~H.~G. Peter, \emph{{Gravitational probes of dark matter
  physics}},  \href{https://arxiv.org/abs/1712.06615}{{\ttfamily 1712.06615}}.

\bibitem{Spergel:1999mh}
D.~N. Spergel and P.~J. Steinhardt, \emph{{Observational evidence for
  selfinteracting cold dark matter}},
  \href{https://doi.org/10.1103/PhysRevLett.84.3760}{\emph{Phys. Rev. Lett.}
  {\bfseries 84} (2000) 3760--3763},
  [\href{https://arxiv.org/abs/astro-ph/9909386}{{\ttfamily
  astro-ph/9909386}}].

\bibitem{Tulin:2017ara}
S.~Tulin and H.-B. Yu, \emph{{Dark Matter Self-interactions and Small Scale
  Structure}}, \href{https://doi.org/10.1016/j.physrep.2017.11.004}{\emph{Phys.
  Rept.} {\bfseries 730} (2018) 1--57},
  [\href{https://arxiv.org/abs/1705.02358}{{\ttfamily 1705.02358}}].

\bibitem{Laha:2016iom}
R.~Laha, \emph{{Effect of hydrodynamical-simulation-inspired dark matter
  velocity profile on directional detection of dark matter}},
  \href{https://doi.org/10.1103/PhysRevD.97.043004}{\emph{Phys. Rev.}
  {\bfseries D97} (2018) 043004},
  [\href{https://arxiv.org/abs/1610.08632}{{\ttfamily 1610.08632}}].

\bibitem{chin2010feshbach}
C.~Chin, R.~Grimm, P.~Julienne and E.~Tiesinga, \emph{Feshbach resonances in
  ultracold gases}, {\emph{Reviews of Modern Physics} {\bfseries 82} (2010)
  1225}, [\href{https://arxiv.org/abs/0812.1496}{{\ttfamily 0812.1496}}].

\bibitem{luo1993weakest}
F.~Luo, G.~C. McBane, G.~Kim, C.~F. Giese and W.~R. Gentry, \emph{The weakest
  bond: Experimental observation of helium dimer}, {\emph{The Journal of
  Chemical Physics} {\bfseries 98} (1993) 3564--3567}.

\bibitem{Rocha:2012jg}
M.~Rocha, A.~H.~G. Peter, J.~S. Bullock, M.~Kaplinghat, S.~Garrison-Kimmel,
  J.~Onorbe et~al., \emph{{Cosmological Simulations with Self-Interacting Dark
  Matter I: Constant Density Cores and Substructure}},
  \href{https://doi.org/10.1093/mnras/sts514}{\emph{Mon. Not. Roy. Astron.
  Soc.} {\bfseries 430} (2013) 81--104},
  [\href{https://arxiv.org/abs/1208.3025}{{\ttfamily 1208.3025}}].

\bibitem{Peter:2012jh}
A.~H.~G. Peter, M.~Rocha, J.~S. Bullock and M.~Kaplinghat, \emph{{Cosmological
  Simulations with Self-Interacting Dark Matter II: Halo Shapes vs.
  Observations}}, \href{https://doi.org/10.1093/mnras/sts535}{\emph{Mon. Not.
  Roy. Astron. Soc.} {\bfseries 430} (2013) 105},
  [\href{https://arxiv.org/abs/1208.3026}{{\ttfamily 1208.3026}}].

\bibitem{Elbert:2016dbb}
O.~D. Elbert, J.~S. Bullock, M.~Kaplinghat, S.~Garrison-Kimmel, A.~S. Graus and
  M.~Rocha, \emph{{A Testable Conspiracy: Simulating Baryonic Effects on
  Self-Interacting Dark Matter Halos}},
  \href{https://doi.org/10.3847/1538-4357/aa9710}{\emph{Astrophys. J.}
  {\bfseries 853} (2018) 109},
  [\href{https://arxiv.org/abs/1609.08626}{{\ttfamily 1609.08626}}].

\bibitem{Kaplinghat:2015aga}
M.~Kaplinghat, S.~Tulin and H.-B. Yu, \emph{{Dark Matter Halos as Particle
  Colliders: Unified Solution to Small-Scale Structure Puzzles from Dwarfs to
  Clusters}}, \href{https://doi.org/10.1103/PhysRevLett.116.041302}{\emph{Phys.
  Rev. Lett.} {\bfseries 116} (2016) 041302},
  [\href{https://arxiv.org/abs/1508.03339}{{\ttfamily 1508.03339}}].

\bibitem{Newman:2012nw}
A.~B. Newman, T.~Treu, R.~S. Ellis and D.~J. Sand, \emph{{The Density Profiles
  of Massive, Relaxed Galaxy Clusters: II. Separating Luminous and Dark Matter
  in Cluster Cores}},
  \href{https://doi.org/10.1088/0004-637X/765/1/25}{\emph{Astrophys. J.}
  {\bfseries 765} (2013) 25},
  [\href{https://arxiv.org/abs/1209.1392}{{\ttfamily 1209.1392}}].

\bibitem{Newman:2012nv}
A.~B. Newman, T.~Treu, R.~S. Ellis, D.~J. Sand, C.~Nipoti, J.~Richard et~al.,
  \emph{{The Density Profiles of Massive, Relaxed Galaxy Clusters: I. The Total
  Density Over 3 Decades in Radius}},
  \href{https://doi.org/10.1088/0004-637X/765/1/24}{\emph{Astrophys. J.}
  {\bfseries 765} (2013) 24},
  [\href{https://arxiv.org/abs/1209.1391}{{\ttfamily 1209.1391}}].

\bibitem{Oh:2010ea}
S.-H. Oh, W.~J.~G. de~Blok, E.~Brinks, F.~Walter and R.~C. Kennicutt, Jr,
  \emph{{Dark and luminous matter in THINGS dwarf galaxies}},
  \href{https://doi.org/10.1088/0004-6256/141/6/193}{\emph{Astron. J.}
  {\bfseries 141} (2011) 193},
  [\href{https://arxiv.org/abs/1011.0899}{{\ttfamily 1011.0899}}].

\bibitem{KuziodeNaray:2007qi}
R.~Kuzio~de Naray, S.~S. McGaugh and W.~J.~G. de~Blok, \emph{{Mass Models for
  Low Surface Brightness Galaxies with High Resolution Optical Velocity
  Fields}}, \href{https://doi.org/10.1086/527543}{\emph{Astrophys. J.}
  {\bfseries 676} (2008) 920--943},
  [\href{https://arxiv.org/abs/0712.0860}{{\ttfamily 0712.0860}}].

\bibitem{Braaten:2017gpq}
E.~Braaten, E.~Johnson and H.~Zhang, \emph{{Zero-range effective field theory
  for resonant wino dark matter. Part I. Framework}},
  \href{https://doi.org/10.1007/JHEP11(2017)108}{\emph{JHEP} {\bfseries 11}
  (2017) 108}, [\href{https://arxiv.org/abs/1706.02253}{{\ttfamily
  1706.02253}}].

\bibitem{Braaten:2017kci}
E.~Braaten, E.~Johnson and H.~Zhang, \emph{{Zero-range effective field theory
  for resonant wino dark matter. Part II. Coulomb resummation}},
  \href{https://doi.org/10.1007/JHEP02(2018)150}{\emph{JHEP} {\bfseries 02}
  (2018) 150}, [\href{https://arxiv.org/abs/1708.07155}{{\ttfamily
  1708.07155}}].

\bibitem{Braaten:2017dwq}
E.~Braaten, E.~Johnson and H.~Zhang, \emph{{Zero-range effective field theory
  for resonant wino dark matter. Part III. Annihilation effects}},
  \href{https://doi.org/10.1007/JHEP05(2018)062}{\emph{JHEP} {\bfseries 05}
  (2018) 062}, [\href{https://arxiv.org/abs/1712.07142}{{\ttfamily
  1712.07142}}].

\bibitem{Clowe:2006eq}
D.~Clowe, M.~Bradac, A.~H. Gonzalez, M.~Markevitch, S.~W. Randall, C.~Jones
  et~al., \emph{{A direct empirical proof of the existence of dark matter}},
  \href{https://doi.org/10.1086/508162}{\emph{Astrophys. J.} {\bfseries 648}
  (2006) L109--L113}, [\href{https://arxiv.org/abs/astro-ph/0608407}{{\ttfamily
  astro-ph/0608407}}].

\bibitem{Springel:2007tu}
V.~Springel and G.~Farrar, \emph{{The Speed of the bullet in the merging galaxy
  cluster 1E0657-56}},
  \href{https://doi.org/10.1111/j.1365-2966.2007.12159.x}{\emph{Mon. Not. Roy.
  Astron. Soc.} {\bfseries 380} (2007) 911--925},
  [\href{https://arxiv.org/abs/astro-ph/0703232}{{\ttfamily
  astro-ph/0703232}}].

\bibitem{Lee:2010hja}
J.~Lee and E.~Komatsu, \emph{{Bullet Cluster: A Challenge to LCDM Cosmology}},
  \href{https://doi.org/10.1088/0004-637X/718/1/60}{\emph{Astrophys. J.}
  {\bfseries 718} (2010) 60--65},
  [\href{https://arxiv.org/abs/1003.0939}{{\ttfamily 1003.0939}}].

\bibitem{Kahlhoefer:2013dca}
F.~Kahlhoefer, K.~Schmidt-Hoberg, M.~T. Frandsen and S.~Sarkar,
  \emph{{Colliding clusters and dark matter self-interactions}},
  \href{https://doi.org/10.1093/mnras/stt2097}{\emph{Mon. Not. Roy. Astron.
  Soc.} {\bfseries 437} (2014) 2865--2881},
  [\href{https://arxiv.org/abs/1308.3419}{{\ttfamily 1308.3419}}].

\bibitem{Lage:2013yxa}
C.~Lage and G.~Farrar, \emph{{Constrained Simulation of the Bullet Cluster}},
  \href{https://doi.org/10.1088/0004-637X/787/2/144}{\emph{Astrophys. J.}
  {\bfseries 787} (2014) 144},
  [\href{https://arxiv.org/abs/1312.0959}{{\ttfamily 1312.0959}}].

\bibitem{Lage:2014yxa}
C.~Lage and G.~R. Farrar, \emph{{The Bullet Cluster is not a Cosmological
  Anomaly}}, \href{https://doi.org/10.1088/1475-7516/2015/02/038}{\emph{JCAP}
  {\bfseries 1502} (2015) 038},
  [\href{https://arxiv.org/abs/1406.6703}{{\ttfamily 1406.6703}}].

\bibitem{Kraljic:2014soa}
D.~Kraljic and S.~Sarkar, \emph{{How rare is the Bullet Cluster (in a
  $\Lambda$CDM universe)?}},
  \href{https://doi.org/10.1088/1475-7516/2015/04/050}{\emph{JCAP} {\bfseries
  1504} (2015) 050}, [\href{https://arxiv.org/abs/1412.7719}{{\ttfamily
  1412.7719}}].

\bibitem{Massey:2017cwf}
D.~Harvey et~al., \emph{{Dark matter dynamics in Abell 3827: new data
  consistent with standard Cold Dark Matter}},
  \href{https://doi.org/10.1093/mnras/sty630}{\emph{Mon. Not. Roy. Astron.
  Soc.} {\bfseries 477} (2018) 669--677},
  [\href{https://arxiv.org/abs/1708.04245}{{\ttfamily 1708.04245}}].

\bibitem{Harvey:2015hha}
D.~Harvey, R.~Massey, T.~Kitching, A.~Taylor and E.~Tittley, \emph{{The
  non-gravitational interactions of dark matter in colliding galaxy clusters}},
  \href{https://doi.org/10.1126/science.1261381}{\emph{Science} {\bfseries 347}
  (2015) 1462--1465}, [\href{https://arxiv.org/abs/1503.07675}{{\ttfamily
  1503.07675}}].

\bibitem{Robertson:2016xjh}
A.~Robertson, R.~Massey and V.~Eke, \emph{{What does the Bullet Cluster tell us
  about self-interacting dark matter?}},
  \href{https://doi.org/10.1093/mnras/stw2670}{\emph{Mon. Not. Roy. Astron.
  Soc.} {\bfseries 465} (2017) 569--587},
  [\href{https://arxiv.org/abs/1605.04307}{{\ttfamily 1605.04307}}].

\bibitem{Read:2018pft}
J.~I. Read, M.~G. Walker and P.~Steger, \emph{{The case for a cold dark matter
  cusp in Draco}}, \href{https://doi.org/10.1093/mnras/sty2286}{\emph{Mon. Not.
  Roy. Astron. Soc.} {\bfseries 481} (2018) 860},
  [\href{https://arxiv.org/abs/1805.06934}{{\ttfamily 1805.06934}}].

\bibitem{Kim:2016ujt}
S.~Y. Kim, A.~H.~G. Peter and D.~Wittman, \emph{{In the Wake of Dark Giants:
  New Signatures of Dark Matter Self Interactions in Equal Mass Mergers of
  Galaxy Clusters}}, \href{https://doi.org/10.1093/mnras/stx896}{\emph{Mon.
  Not. Roy. Astron. Soc.} {\bfseries 469} (2017) 1414--1444},
  [\href{https://arxiv.org/abs/1608.08630}{{\ttfamily 1608.08630}}].

\bibitem{efimov1970energy}
V.~Efimov, \emph{Energy levels arising from resonant two-body forces in a
  three-body system}, {\emph{Physics Letters B} {\bfseries 33} (1970)
  563--564}.

\bibitem{efimov1973energy}
V.~Efimov, \emph{Energy levels of three resonantly interacting particles},
  {\emph{Nuclear Physics A} {\bfseries 210} (1973) 157--188}.

\bibitem{2008PhRvL.100n0404G}
A.~O. {Gogolin}, C.~{Mora} and R.~{Egger}, \emph{{Analytical Solution of the
  Bosonic Three-Body Problem}},
  \href{https://doi.org/10.1103/PhysRevLett.100.140404}{\emph{Physical Review
  Letters} {\bfseries 100} (2008) 140404},
  [\href{https://arxiv.org/abs/0802.0549}{{\ttfamily 0802.0549}}].

\bibitem{1994Sci...266.1345S}
W.~{Schollkopf} and J.~P. {Toennies}, \emph{{Nondestructive Mass Selection of
  Small van der Waals Clusters}},
  \href{https://doi.org/10.1126/science.266.5189.1345}{\emph{Science}
  {\bfseries 266} (1994) 1345--1348}.

\bibitem{voigtsberger2014imaging}
J.~Voigtsberger, S.~Zeller, J.~Becht, N.~Neumann, F.~Sturm, H.-K. Kim et~al.,
  \emph{Imaging the structure of the trimer systems $^4\!\text{He}_3$ and
  $^3\!\text{He}$\, $^4\!\text{He}_2$}, {\emph{Nature Communications}
  {\bfseries 5} (2014) 5765}.

\bibitem{kraemer2006evidence}
T.~Kraemer, M.~Mark, P.~Waldburger, J.~Danzl, C.~Chin, B.~Engeser et~al.,
  \emph{Evidence for {E}fimov quantum states in an ultracold gas of caesium
  atoms}, {\emph{Nature} {\bfseries 440} (2006) 315}.

\bibitem{Braaten:2008kx}
E.~Braaten, H.~W. Hammer, D.~Kang and L.~Platter, \emph{{Three-Body
  Recombination of Identical Bosons with a Large Positive Scattering Length at
  Nonzero Temperature}},
  \href{https://doi.org/10.1103/PhysRevA.78.043605}{\emph{Phys. Rev.}
  {\bfseries A78} (2008) 043605},
  [\href{https://arxiv.org/abs/0801.1732}{{\ttfamily 0801.1732}}].

\bibitem{macek2006exact}
J.~Macek, S.~Y. Ovchinnikov and G.~Gasaneo, \emph{Exact solution for three
  particles interacting via zero-range potentials}, {\emph{Physical Review A}
  {\bfseries 73} (2006) 032704}.

\bibitem{Platter:2004he}
L.~Platter, H.~W. Hammer and U.-G. Meissner, \emph{{The Four boson system with
  short range interactions}},
  \href{https://doi.org/10.1103/PhysRevA.70.052101}{\emph{Phys. Rev.}
  {\bfseries A70} (2004) 052101},
  [\href{https://arxiv.org/abs/cond-mat/0404313}{{\ttfamily
  cond-mat/0404313}}].

\bibitem{Hammer:2006ct}
H.~W. Hammer and L.~Platter, \emph{{Universal Properties of the Four-Body
  System with Large Scattering Length}},
  \href{https://doi.org/10.1140/epja/i2006-10301-8}{\emph{Eur. Phys. J.}
  {\bfseries A32} (2007) 113--120},
  [\href{https://arxiv.org/abs/nucl-th/0610105}{{\ttfamily nucl-th/0610105}}].

\bibitem{2008arXiv0810.3876V}
J.~{von Stecher}, J.~P. {D'Incao} and C.~H. {Greene}, \emph{{Four-body legacy
  of the Efimov effect}}, \href{https://doi.org/10.1038/nphys1253}{\emph{Nature
  Physics} {\bfseries 5} (2009) 417--421},
  [\href{https://arxiv.org/abs/0810.3876}{{\ttfamily 0810.3876}}].

\bibitem{2009PhRvL.102n0401F}
F.~{Ferlaino}, S.~{Knoop}, M.~{Berninger}, W.~{Harm}, J.~P. {D'Incao}, H.-C.
  {N{\"a}gerl} et~al., \emph{{Evidence for Universal Four-Body States Tied to
  an Efimov Trimer}},
  \href{https://doi.org/10.1103/PhysRevLett.102.140401}{\emph{Physical Review
  Letters} {\bfseries 102} (2009) 140401},
  [\href{https://arxiv.org/abs/0903.1276}{{\ttfamily 0903.1276}}].

\bibitem{2011PhRvL.107t0402V}
J.~{von Stecher}, \emph{{Five- and Six-Body Resonances Tied to an Efimov
  Trimer}},
  \href{https://doi.org/10.1103/PhysRevLett.107.200402}{\emph{Physical Review
  Letters} {\bfseries 107} (2011) 200402},
  [\href{https://arxiv.org/abs/1106.2319}{{\ttfamily 1106.2319}}].

\bibitem{Patrignani:2016xqp}
{\scshape Particle Data Group} collaboration, C.~Patrignani et~al.,
  \emph{{Review of Particle Physics}},
  \href{https://doi.org/10.1088/1674-1137/40/10/100001}{\emph{Chin. Phys.}
  {\bfseries C40} (2016) 100001}.

\bibitem{Armendariz-Picon:2013jej}
C.~Armendariz-Picon and J.~T. Neelakanta, \emph{{How Cold is Cold Dark
  Matter?}}, \href{https://doi.org/10.1088/1475-7516/2014/03/049}{\emph{JCAP}
  {\bfseries 1403} (2014) 049},
  [\href{https://arxiv.org/abs/1309.6971}{{\ttfamily 1309.6971}}].

\bibitem{Steigman:2012nb}
G.~Steigman, B.~Dasgupta and J.~F. Beacom, \emph{{Precise Relic WIMP Abundance
  and its Impact on Searches for Dark Matter Annihilation}},
  \href{https://doi.org/10.1103/PhysRevD.86.023506}{\emph{Phys. Rev.}
  {\bfseries D86} (2012) 023506},
  [\href{https://arxiv.org/abs/1204.3622}{{\ttfamily 1204.3622}}].

\bibitem{Binder:2017rgn}
T.~Binder, T.~Bringmann, M.~Gustafsson and A.~Hryczuk, \emph{{Early kinetic
  decoupling of dark matter: when the standard way of calculating the thermal
  relic density fails}},
  \href{https://doi.org/10.1103/PhysRevD.96.115010}{\emph{Phys. Rev.}
  {\bfseries D96} (2017) 115010},
  [\href{https://arxiv.org/abs/1706.07433}{{\ttfamily 1706.07433}}].

\bibitem{Kaplinghat:2013xca}
M.~Kaplinghat, R.~E. Keeley, T.~Linden and H.-B. Yu, \emph{{Tying Dark Matter
  to Baryons with Self-interactions}},
  \href{https://doi.org/10.1103/PhysRevLett.113.021302}{\emph{Phys. Rev. Lett.}
  {\bfseries 113} (2014) 021302},
  [\href{https://arxiv.org/abs/1311.6524}{{\ttfamily 1311.6524}}].

\bibitem{Binder:2017lkj}
T.~Binder, M.~Gustafsson, A.~Kamada, S.~M.~R. Sandner and M.~Wiesner,
  \emph{{Reannihilation of self-interacting dark matter}},
  \href{https://doi.org/10.1103/PhysRevD.97.123004}{\emph{Phys. Rev.}
  {\bfseries D97} (2018) 123004},
  [\href{https://arxiv.org/abs/1712.01246}{{\ttfamily 1712.01246}}].

\end{thebibliography}\endgroup

\end{document}